\documentclass[final,3p,times]{elsarticle}

\usepackage{epsfig}

\usepackage{amssymb,amsmath }

\usepackage{hyperref}
\hypersetup{colorlinks=true, urlcolor=blue}

\usepackage{subfigure}
\usepackage{booktabs}

\fussy

\newcommand{\ud}{\mathrm{d}}
\newcommand{\y}{y}

\newcommand{\Halmos}{\mbox{\quad$\square$}}

\newtheorem{proposition}{Proposition}

\newtheorem{definition}{Definition}
\newtheorem{assumption}{Assumption}
\newproof{proof}{{\em Proof}}
\newproof{pot}{{\em Proof of Theorem \ref{thm2}}}

\journal{}

\begin{document}

\begin{frontmatter}


\title{Stock Price Dynamics and Option Valuations under Volatility Feedback Effect}


\author[IM]{Juho Kanniainen\corref{cor1}}
\ead{juho.kanniainen@tut.fi}
\author[MAT]{Robert Pich\'e}
\ead{robert.piche@tut.fi}
\cortext[cor1]{Corresponding author. Telephone +358 40 707 4532. Fax +358 3 3115 2027.}
\address[IM]{Tampere University of Technology, Department of Industrial Management.\\ P.O. Box 541, FI-33101 Tampere, Finland.}
\address[MAT]{Tampere University of Technology, Department of Mathematics.\\ P.O. Box 553, FI-33101 Tampere, Finland.}
%


\begin{abstract}
{ 
According to the volatility feedback effect, an unexpected increase in squared volatility leads to an immediate decline in the price-dividend ratio. In this paper, we consider the properties of stock price dynamics and option valuations under the volatility feedback effect by modeling the joint dynamics of stock price, dividends, and volatility in continuous time. Most importantly, our model predicts the negative effect of an increase in squared return volatility on the value of deep-in-the-money call options and, furthermore, attempts to explain the volatility puzzle. We theoretically demonstrate a mechanism by which the market price of diffusion return risk, or an equity risk-premium, affects option prices and empirically illustrate how to identify that mechanism using forward-looking information on option contracts. Our theoretical and empirical results support the relevance of the volatility feedback effect. Overall,  the results indicate that the prevailing practice of ignoring the time-varying dividend yield in option pricing can lead to oversimplification of the stock market dynamics.
}

\end{abstract}


\end{frontmatter}







\section{Introduction}

The fundamental importance of time varying volatility has {  long been} recognized in statistical finance and financial economics, and many scientific findings thereof have been well accepted and exploited in these disciplines. The topic is motivated by strong evidence that volatility does not remain constant over time. Recently, in October 2008, the volatility index of S\&P 500, VIX, hit 80\% whereas its average (1990--2009) was about 20\%. Empirically, also the price-dividend ratio (or its reciprocal, the dividend yield) is time-varying and one of the 'stylized facts' of financial markets is that changes in the price-dividend ratio are negatively correlated with volatility. {  Many theories, ofi which the so-called volatility feedback effect (sometimes called the risk-premium effect) is one,} explain the empirically observed negative correlation between volatility and stock price  \citep*[see, e.g.,][]{Black76,Merton80,Pindyck84,French87,Campbell92,Bekaert00,Raberto05,Bollerslev06}. According to the theory of volatility feedback effect, an unexpected increase in squared volatility leads to an immediate decline in the stock price, because cash flows are discounted at a higher rate. Therefore, an exogenous increase in squared volatility generates additional return volatility as stock prices respond and adjust to new information about the cost of capital. In addition, the relation between volatility and returns can { (at least partly)} be explained by the leverage effect, which extends from changes in the firm's value to changes in stock returns and volatility. The difference lies in causality -- the volatility feedback effect theory contends that changes in volatility may produce return shocks, whereas the leverage hypothesis predicts that return shocks lead to changes in volatility. Also the leverage effect is widely examined in the literature \citep*[see, e.g.,][and references therein]{Florescu09}. 

The time-varying price-dividend ratio (or the dividend yield) and its relation to stochastic volatility is well documented in the empirical literature, but the current option pricing literature does not sufficiently characterize the joint dynamics of dividends, volatility, and stock price; instead, typically in option pricing, dividends are either ignored or the dividend yield is assumed to be constant at best. In this paper, we aim to show that the prevailing practice of ignoring the modelling of the joint dynamics of dividends, volatility, and stock prices {  is} inconsistent not only with respect to financial data but also with respect to financial theory itself. This oversimplification can lead to mispricing of options and a misestimation of the effects of the return risk and volatility risk on option values. Our main goals are as follows:
\begin{itemize}
\item Model the joint stochastic dynamics of return volatility, dividends, and stock price with volatility feedback in continuous time by determining the underlying stock as a claim for future random dividends with a stochastic discount rate.
\item Express the relation between dividend growth volatility and return volatility and {  solve the volatility puzzle (i.e. return volatility is too high compared to dividend growth volatility).}
\item Show that the correlation between returns and volatility can be divided into two components: leverage effect and volatility feedback effect.
\item Demonstrate a {  mechanism by} which the market price of return risk, or equity risk-premium, affects option prices.
\item {  Show that, contrary to the prevailing view,} an increase in squared return volatility can {\em negatively} affect the price of deep-in-the-money call options.
\item Illustrate how to obtain forward-looking estimates for the price of diffusion return risk using information on option contracts.
\end{itemize}

One of the main {  implications} of Black-Scholes theory is the irrelevancy of the equity risk premium in option valuation (i.e. option values are not functions of the expected rate of return). We, however, aim to show that the price of return risk determines the sensitivity of the dividend yield to return volatility and thereby affects option valuations, and consequently, the market price of return risk is needed as an input to price options under our framework. This, on the other hand, allows us to produce forward-looking option-implied estimates for the market price of the diffusion return risk and the volatility risk premium as a part of the calibration procedure of our model. These option-based estimates can be obtained using option data alone without the need of historical stock price data, which is in contrast to the traditional literature that usually uses a series of equity market indices producing backward-looking estimates for the market price of diffusion return risk. 

In the early literature, \citet*{Carr08} {  provided} a welcome exception by {  proposing} a model that aims to capture the volatility feedback effect and estimate the jump risk and the variance rate risk using option data. \citet{Bakshi10} specified a model to estimate market prices of different sources of risks using information on both time-series returns and options prices. However, {  the approaches in these two papers differ} markedly from ours. Most importantly, they assume a constant dividend yield, an assumption that contradicts the empirical evidence of varying price-dividend ratio and the theory of the volatility feedback effect. Moreover, \citet*{Carr08} try to capture the volatility feedback effect directly by assuming a negative statistic correlation between business risk and stock price {  without modeling the changes that the underlying asset price undergoes in volatility.} In addition, in contrast to our paper, the price of the {\em diffusion} return risk does not appear in option pricing formulas under the risk-neutral measure in \citep*{Carr08,Bakshi10}. Our paper is also related to \citep{Kanniainen11}, which integrates the stochastic dynamics of interest rates, dividends, and stock prices and valuates options accordingly. Despite of some methodological similarities between the papers, \citep{Kanniainen11} differs substantially from the present paper; whereas \citet*{Kanniainen11} focuses on the joint dynamics of spot rate and dividends but ignores the volatility feedback effect, in this paper we investigate the stock market dynamics and options prices under the volatility feedback but, for simplicity, assume constant interest rates. 

The paper is organized as follows. In Section \ref{SEC:Stock_Market_Dynamics}, we present our model setup, solve the price-dividend ratio with it, and {  study} stock market dynamics under our assumptions. In Section \ref{SEC:Option_Pricing}, we show how to price options under our settings, and in Section \ref{SEC:Empirical_Analysis} we provide an empirical illustration. The final section discusses the results and draws conclusions.

\section{Stock market dynamics}\label{SEC:Stock_Market_Dynamics}
\subsection{Model setup}
Let $\{P_t; t\geq 0\}$ denote the stock price process and $\{D_t; t\geq 0\}$ the instantaneous dividend stream and let us assume that both $\{P_t; t\geq 0\}$ and $\{D_t; t\geq 0\}$ evolve on $\mathbb{R}_+$. We define the cumulative stock returns as follows:
\begin{definition}\label{DEF:dR}{\bf Cumulative stock return.}
  The cumulative stock return from dividends and changes in prices satisfies
  \[
  \ud R_t = \frac{\ud P_t + D_t \ud t}{P_t}.
  \]
\end{definition}
Thus $\{R_t; t \geq 0\}$ represents the instantaneous total return including price appreciations and dividends. To focus on the
characterization of stock market dynamics and valuation of options with volatility feedback condition, and to maintain conciseness and readability, we employ pure diffusion-based models and {  leave} extensions, including jumps and non-affine volatility models, for future research. In the following, we characterize the dynamics of cumulative stock returns and return volatility.

\begin{assumption}\label{ASS:dR,dx}
The cumulative stock return and its volatility evolve stochastically as
\begin{align}
\ud R_t &= \left(r + \gamma x_t^2\right) \ud t + x_t \ud B_t^r\label{EQ:dR}\\
\ud x_t &= -\beta x_t \ud t + \sigma_x \ud B_t^x,\label{EQ:dx}
\end{align}
  where $x_t$ represents instantaneous return volatility, and $r, \gamma, \beta$, and $\sigma_x$ are constant positive real numbers. Moreover, $B^r$ and $B^x$ are Brownian motions, $\ud B_t^r \ud B_t^x = \rho_{rx,t} \ud t$, and $x_0 := x$, $x \in \mathbb{R}$.
\end{assumption}

This assumption consists of two parts. First, in Eq.\ (\ref{EQ:dR}) we assume that the expected rate of return (including both price appreciation and dividend yield) depends on squared return volatility and results in an ICAPM type equilibrium \citep*[see][]{Merton73,Merton80}, where  $r$ denotes the risk-free interest rate and where, under certain assumptions, the price of diffusion return risk, $\gamma$, represents the coefficient of relative risk aversion \citep{Merton80,Campbell93}. This classical risk-return tradeoff relation is widely used in various contexts in the financial literature with time variation in second moments. Second, we follow \citet{Heston93} and assume that volatility follows an Ornstein-Uhlenbeck process according to Eq. (\ref{EQ:dx}). In fact, this volatility model in Heston's paper \citep[see p. 328, Eq. (2) therein]{Heston93} {  has its roots in the Stein-Stein model} \citep{Stein91}. As Heston shows, if volatility follows Eq. (\ref{EQ:dx}), then squared volatility, $h_t = x_t^2$, follows the squared root process: 
\begin{equation}\label{EQ:dh}
\ud h_t = \kappa(\theta - h_t)\ud t + \sigma_h \sqrt{h_t}\ud B_t^x,
\end{equation}
where $\kappa = 2 \beta$, $\theta = \sigma_x^2/(2 \beta)$, and $\sigma_h = 2\sigma_x$. This affine model is arguably among the most widely used continuous-time stochastic volatility models in finance. Notice that in (\ref{EQ:dh}) the correlation between returns and volatility, $\rho_{rx,t}$, is assumed to be time-varying, and as seen later, can be endogenously determined by specifying the process for the dividend stream and by applying the transversality condition. Originally, and typically, the correlation is assumed to be constant in time.\footnote{{ Notice that the sign of $x$ is irrelevant.} It is the squared volatility that matters: the statistical properties of the stock return dynamics, including the correlation between squared volatility and returns, are the same with $x_t$ and $-x_t$. }

Second, we assume that the stock (the stock index) pays dividends continuously with stochastic dividend growth volatility. In particular, similarly to \citep{Brennan01} and \citep{Ang07}, instantaneous dividends are assumed to follow a geometric process.

\begin{assumption}\label{ASS:dD}
(i) The stochastic differential of dividends is given by
\[
\ud D_t = \alpha D_t \ud t + \y_t D_t \ud B^d_t
\]
with $\ud B^d_t \ud B^x_t = \rho_{dx} \ud t$, where the correlation coefficient of dividend growth and return volatility, $\rho_{dx} \in [-1,1]$, and the expected rate of dividend growth, $\alpha \in \mathbb{R}$, are constant, and where dividend growth volatility, $\y_t$, is stochastic. Moreover, $D_0 := D$, $D>0$.\\

(ii) The covariance and correlation between dividends and return volatility {  are related according to}
\[
  \mathrm{sign}\left(\frac{\ud }{\ud \tau}\mathrm{Cov}_t \left(D_t, (x_t^2 \right))|_{\tau = t} \right) = \mathrm{sign}(\rho_{dx}).
  \]
\end{assumption}

In assumption ({\em i}), a negative correlation between dividends and return volatility can be thought of as representing the leverage effect: the greater ({  resp.} less) the dividends, the greater ({  resp.} less) the stock price, and because of financial leverage, the less ({  resp.} greater) the return volatility. The assumption of lognormal instantaneous dividends (cash flows) is quite common in the literature \citep*[see, e.g.,][]{McDonald85,Brennan01,Bakshi05,Ang07,Kanniainen09b,Kanniainen11}. Assumption ({\em ii}) implies that {  the correlation coefficient $\rho_{dx}$ has the same sign as} the covariance between dividends and squared return volatility. Consequently, because $\rho_{dx}$ is assumed to be a constant, the sign of the covariance between dividends and squared return volatility does not change over time. Note that the standard Heston model implies the same relation for the covariance between returns and squared return volatility and the corresponding correlation coefficient.

Together Assumptions \ref{ASS:dR,dx} and \ref{ASS:dD} imply that the price process is given by
\[\begin{split}
  \ud P_t &= \left(r + \gamma x_t^2 \right)P_t \ud t - D_t \ud t + x_t P_t \ud B_t^r\\
  &= \left(r + \gamma x_t^2 - \delta_t \right)P_t \ud t + x_t P_t \ud B_t^r,
\end{split}\]
where $\delta_t:= D_t / P_t$ represents the time-varying instantaneous dividend yield. 

\begin{assumption}\label{ASS:transversality}
  By assuming transversality, we express the stock price, $p(D_t, x_t)$, $P_t = p(D_t, x_t)$, as the expected value of discounted dividends, conditional upon the present information:
  \begin{equation}\begin{split}\label{EQ:p}
    p(D_t, x_t) &= \mathbb{E}_{D,x}\int_t^{\infty} \exp \left[-\int_t^s \left(r + \gamma x_u^2 \right) \ud u \right] D_s \ud s\\
    &= D_t \times \mathbb{E}_x \int_t^{\infty} \exp \bigg[\int_t^s \left(\alpha - r - \gamma x_u^2 - \frac{1}{2} \y_u^2 \right) \ud u + \int_t^s \y_u \ud B_u^d \bigg] \ud s < \infty.
  \end{split}\end{equation}
  Here $r + \gamma x_u^2$ represents the instantaneous stochastic cost of capital at time $u$.
\end{assumption}

The above expression clarifies the concept of the volatility feedback effect, according to which the stock price is determined by the expected value of discounted dividends, where the cost of capital depends positively on squared return volatility. As squared spot volatility increases, then also the future values of return volatility are expected to increase, and  future dividends are thereby discounted at a higher rate. Then according to Eq.\ (\ref{EQ:p}), the current stock price immediately responds negatively to an increased cost of capital, generating additional return volatility as stock prices adjust to new information \citep*[see, e.g.,][]{Pindyck84,French87,Campbell92,Bekaert00}.

In contrast to our characterization, the existing literature seeks to capture the volatility feedback effect differently and, in fact, in numerous ways. Typically, models are in discrete-time and based on GARCH type settings \citep*[see, e.g.,][]{Campbell92,Bekaert00}{ ; continuous-time characterizations have also been proposed \citep*[see, e.g.,][]{Bollerslev06,Carr08}.} More importantly, many papers in the volatility feedback literature assume that the risk-premium depends on dividend growth volatility instead of return volatility \citep*[see, e.g.,][and the references therein]{Campbell92,Wu01}, but, on the other hand, in the conditional CAPM literature it is the conventional ICAPM type risk-return trade-off with a linear relation between risk-premium and squared return volatility that has been the primary target of investigation. Empirically, stock return volatility is admittedly much higher than dividend growth volatility, a phenomenon extensively investigated in the literature, and hence {  estimates} of the risk-return trade-off parameter can {  differ} depending on whether the risk premium is assumed to depend on return volatility or dividend growth volatility. In addition, in several papers, it makes sense to work within the log-linear approximate asset pricing framework of \citet*{Campbell88} \citep*[see, e.g.,][]{Campbell92,Wu01}, but we show it more worthwhile under our settings to use numerical methods to solve the price-dividend ratio.

Under these assumptions, we aim to investigate the joint dynamics of and relation between stock price, return volatility, and dividend stream. In particular, we seek to determine the relations of return volatility to dividend growth volatility and to the price-dividend ratio (and its reciprocal, the dividend yield). 

\begin{definition}\label{DEF:f}{\bf Price-Dividend Ratio.}
  $f:\mathbb{R} \to \mathbb{R}_+$ denotes the price-dividend ratio and satisfies $p(D_t, x_t) = D_t f(x_t)$.
\end{definition}

We can see from Eq.\ (\ref{EQ:p}) that for all $x > 0$, $p(D, x) = p(D, -x)$ and so $f(x) = f(-x)$. Practically, this holds because the stochastic cost of capital is determined by squared volatility, $x^2$, and hence the sign of volatility, $x$, does not affect the stock price dynamics. Thus we can deduce that $f$ is an even function, i.e., $f(x) = f(-x), f_x(x) = -f_x(-x)$, and $f_{xx}(x) = f_{xx}(-x)$ for all $x> 0$, where $f_x$ and $f_{xx}$ denote first and second order derivatives. Moreover, for the stock price to be a {  continuously} differentiable function, we can impose that $f_x(0) = 0$. We will use this property as a boundary condition to solve the price-dividend ratio for $x \geq 0$ .  

\subsection{Solution}

We assume that dividend growth volatility is stochastic and interlinked with stochastic return volatility. In fact, it is easy to show that if dividend growth is assumed to be IID, i.e., $\y_t$ is constant over time, then for $|\rho_{dx}| \neq 1$ the stock price is not real-valued for all $x \geq 0$ under our assumptions. This is intuitive, because with constant dividend growth volatility the only source of stochasticity in return volatility would be the stochastic cost of capital, which, according to CAPM, is determined by return volatility itself. On the whole, because the assumption of IID dividend growth is not reasonable under our settings, the rest of the study considers dividend growth volatility as an endogenously determined time-varying and stochastic variable. We { offer} now a solution to dividend growth volatility as a function of return volatility. Using this solution, we then present a non-homogeneous ordinary differential equation that the price-dividend ratio must satisfy. 

\begin{proposition}\label{PROP:y}
Suppose that the above assumptions hold. Then {  the} dividend growth volatility, $\y(x)$, $\y_t = \y(x_t)$, satisfies
  \begin{equation}\label{EQ:y(x)}
  \y(x) = -\rho_{dx} \sigma_x \frac{f_x(x)}{f(x)} + \mathrm{sign}(x) \sqrt{x^2 - (1 - \rho_{dx}^2) \left(\sigma_x \frac{f_x(x)}{f(x)} \right)^2}.
\end{equation}
\end{proposition}

\proof{}
  By applying It\^o's Lemma {  with} Definition \ref{DEF:f} and noting that $\partial p(x,D)/\partial D = f(x)$, the price process can be expressed as
  \begin{equation}\begin{split}\label{EQ:dP}
    \ud P_t &= \left(\alpha f(x_t) D_t -\beta x_t f_x(x_t) D_t + f_x(x_t) \rho_{dx} \sigma_x y_t D_t + \frac{1}{2} f_{xx}(x_t) \sigma_x^2  D_t \right) \ud t\\
    &+ f_x(x_t) \sigma_x D_t \ud B_t^x + f(x_t) \y_t D_t \ud B_t^d\\
    &= \left(\alpha + \left(\rho_{dx} \sigma_x y_t - \beta x_t \right)\frac{f_x(x_t)}{f(x_t)} + \frac{1}{2} \sigma_x^2 \frac{f_{xx}(x_t)}{f(x_t)}  \right) P_t \ud t + \sigma_x \frac{f_x(x_t)}{f(x_t)} P_t \ud B_t^x + \y_t P_t \ud B_t^d.
  \end{split}\end{equation}
The above reasoning is also applied, e.g., in \citep{Brennan01} and \citep{Ang07}. From the relation between the return process, volatility process, and dividend process we obtain with $\y_t = \y(x_t)$ that
  \[
  x^2 = \y(x)^2 + \sigma_{x}^2 \frac{f_x(x)^2}{f(x)^2} + 2 \rho_{dx} \y(x) \sigma_x \frac{f_x(x)}{f(x)}.
  \]
  When we solve the above with respect to $\y(x)$, we obtain
  \begin{equation}\label{EQ:y(x)_proof}
  \y(x) = -\rho_{dx} \sigma_x \frac{f_x(x)}{f(x)} \pm \sqrt{x_t^2 - (1 - \rho_{dx}^2) \left(\sigma_x \frac{f_x(x)}{f(x)} \right)^2}.
  \end{equation}
  First, {  if $x>0$, we} choose the greater of the two roots above. To understand this, note that by It\^o's lemma
  \[
  \frac{\ud }{\ud \tau}\mathrm{Cov}_t \left(D_t, (x_t^2 )\right)|_{\tau = t} = 2 \sigma_x x_t \y(x_t) D_t \rho_{dx},
  \]
  which implies under Assumption \ref{ASS:dD} ({\em ii}) that $\mathrm{sign}\left(\y(x)\right) = \mathrm{sign}(x)$; hence we must choose the greater of the two roots to have a strictly positive $\y$ for any $x>0$. At this point, note also that if $\rho_{dx} < 0$, then the first term on the right hand side of Eq.\ (\ref{EQ:y(x)_proof}) is strictly negative for any $x>0$. This implies that under certain conditions, the structural parameters are bounded. The above also implies that $\y(x) < 0$ for $x<0$; hence we always choose the smaller of the roots for any $x<0$.\Halmos \\
\endproof

Note that the above Proposition implies that dividend growth volatility $\y(x) = 0$ if and only if return volatility $x=0$. Furthermore, dividend growth volatility $\y(x) > 0$ and $\y(-x) = -\y(x)$ for any $x>0$. It also follows that under our assumptions the price of the stock {  would be difficult to solve} by Monte Carlo methods. In particular, to compute Eq.\ (\ref{ASS:transversality}), we must solve $y_u = \y(x_u)$ iteratively, which is determined by the price-dividend ratio, $f(x)$, which again directly determines the stock price. Therefore, we must look for a solution in another direction.

Using the solution for dividend growth volatility, we can formulate a differential equation that the price-dividend ratio must satisfy:

\begin{proposition}\label{PROP:DE}
  Suppose that the conditions of Proposition \ref{PROP:y} hold. Then the price-dividend ratio satisfies the following relation:
  \begin{equation}\label{EQ:DE}
    (\y(x) \sigma_x \rho_{dx} -\beta x )f_x (x) + \frac{1}{2} \sigma_x^2 f_{xx}(x) - (r + \gamma x^2 - \alpha)f(x) = - 1,
  \end{equation}
  where $y(x)$ is given in Equation (\ref{EQ:y(x)}). For the interval $x \geq 0$, the boundary conditions are $f(x) = 0 \text{\ as \ } x \to \infty$ and $f_x(x) = 0 \text{\ at \ } x = 0$.
\end{proposition}
 
\proof{} The result is directly obtained by matching the drift term of the stock price process (\ref{EQ:dP}) and the assumed required rate of return minus the dividend yield, $r + \gamma x^2 - 1/f(x)$. 

The first boundary condition is based on the fact that future dividends are discounted at an extremely high cost of capital if the volatility of stock returns is extremely high. The second condition is imposed by the differentiability of $f$ at $x=0$.\Halmos \\
\endproof

This model is hardly tractable analytically, especially with $\rho_{dx} \neq 0$; therefore, we employ numerical methods to solve it.\footnote{With $\rho_{dx}=0$, Eq.\ (\ref{EQ:DE}) is a non-homogeneous ODE, whose associated homogeneous equation belongs to the class of  degenerate hypergeometric equations. However, the solution is very complicated and extremely hard to interpret \citep*[see][]{Campbell02,Polyanin03}.} The numerical solution is described in the Appendix.

In Figure \ref{FIG:f(x)}, we have {  plotted} the price-dividend ratio as a function of return volatility for $x \geq 0$. We do not show the corresponding plot for the interval of $x \leq 0$ because it can be considered as a "mirror image" of Figure \ref{FIG:f(x)} as $f(-x) = f(x)$ ($f$ is an {  even} function). \citet{Bakshi97} estimated from option prices that the volatility of the variance, $\sigma_h$ is about 0.4, depending on the moneyness of the options, implying that $\sigma_x = \sigma_h/2 \approx 0.2$. Moreover, the speed of adjustment of the squared volatility process, $\kappa$, almost equalled one, and thus $\beta = \kappa/2 \approx 0.5$. We also note that these values are consistent with their estimation of {  the long-run average squared volatility (variance)}, $\theta \approx 0.04$. In addition, we assume that the instantaneous risk-free interest {  rate is} $r = 0.02$ and the correlation between the returns and dividend growth {  is} $\rho_{dx} = -0.5$. {  In Figure \ref{FIG:f(x)} (a), we have plotted three curves by varying the price of diffusion return risk, $\gamma$, the expected dividend growth rate, $\alpha$, and the correlation between return volatility and dividend growth, $\rho_{dx}$. In Figure \ref{FIG:f(x)} (b), we perform a sensitivity analysis by varying $\beta$ and $\sigma_x$ from Bakshi's estimates \cite{Bakshi97}.}

\begin{figure}[h]
\begin{center}
  \includegraphics*[width=0.7\textwidth]{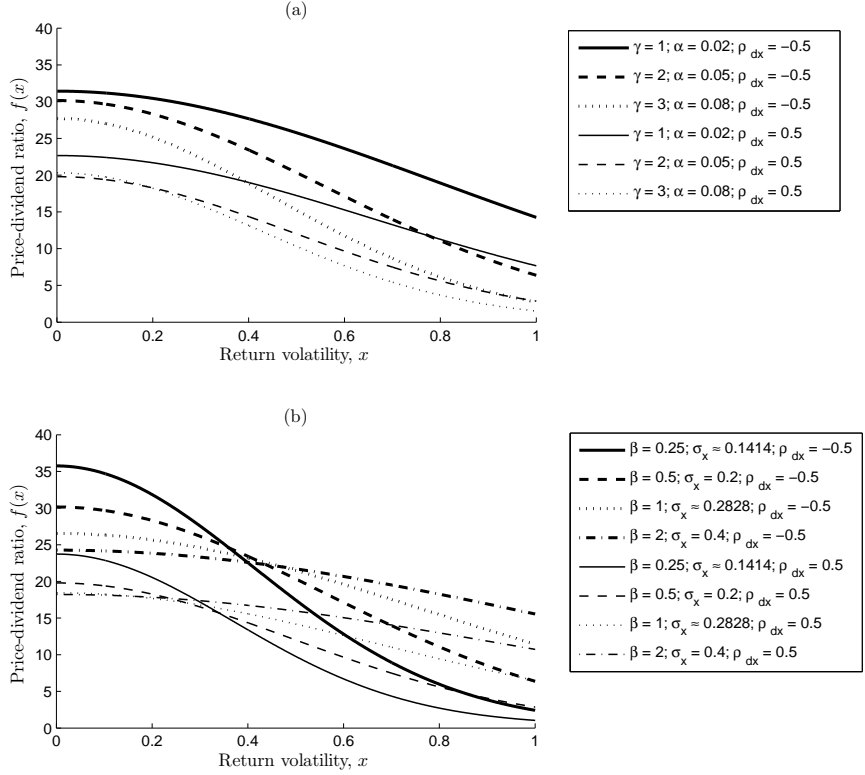}
\caption{ \footnotesize{  The price-dividend ratio, $f(x)$, with respect to return volatility, $x$. In plot (a), parameters are $r = 0.02$, $\beta = 0.5$, $\sigma_x = 0.2$, $\rho_{dx} = -0.5$ (thicker lines) or $\rho_{dx} = 0.5$ (thinner lines), and $\gamma$ and $\alpha$ vary. In plot (b), parameters are $r = 0.02$, $\gamma = 2$, $\alpha = 0.05$, $\rho_{dx} = -0.5$ (thicker lines) or $\rho_{dx} = 0.5$ (thinner lines), and $\beta$ and $\sigma_x$ vary. In particular, the relation between $\beta$ and $\sigma_x$ is determined with $\sigma_x = \sqrt{2\theta\beta}$, where the average variance $\theta = 0.04$.}}
\label{FIG:f(x)}
\end{center}
\end{figure}

As expected, the stock price is a monotone decreasing function of squared return volatility. {  At the origin}, the first derivative is zero, as required by the condition of $f_x(0) = 0$. Note how a greater  $\gamma$ implies a lower stock price even with a greater expected dividend growth rate. This seems reasonable, because an increase in $\gamma$ increases the stochastic risk premium, $\gamma x^2$. Note that if $\gamma$ were equal to zero, then Assumption \ref{ASS:transversality} would  imply that $\alpha$ should be less than the risk-free interest rate; otherwise the stock price would not be well defined. In fact, a greater stochastic risk premium implies a greater upper bound of the expected dividend growth. For example, with $\gamma = 1$ and $\alpha = 0.08$, the {  numerical solver\footnote{{ \textsc{Matlab} \texttt{bvp4c}}} } returns an error message and we cannot find a suitable numerical solution for the price-dividend ratio because dividend growth would then be too high compared to the (stochastic) risk premium, but with $\gamma = 3$ and $\alpha = 0.08$, a solution does exist (see Figure \ref{FIG:f(x)}). 

As the above analysis shows, the price-dividend ratio, $f_t = f(x_t)$, varies as return volatility changes; hence the dividend yield, $\delta_t \equiv 1/f_t$, depends on return volatility and is stochastic. Moreover, the correlation between returns and volatility becomes stochastic under our settings:

\begin{proposition}
  Suppose that the conditions of Proposition \ref{PROP:y} hold. Then the correlation between returns and return volatility, $\rho_{rx,t} = \rho_{rx}(x_t)$,
can be expressed as
    \begin{equation}\label{EQ:rho_rx}
    \rho_{rx}(x) = \mathrm{sign}(x) \frac{\sigma_x \frac{f_x(x)}{f(x)} + \y(x) \rho_{dx}}{\sqrt{\sigma_x^2 \frac{f_x(x)^2}{f(x)^2} + \y(x)^2 + 2 \rho_{dx} \sigma_x \frac{f_x(x)}{f(x)} \y(x)}},
    \end{equation}
    where $y(x)$ is given in Equation (\ref{EQ:y(x)}) and $f$ satisfies Eq.\ (\ref{EQ:DE}). Note that $\rho_{rx}(-x) = \rho_{rx}(x)$.
\end{proposition}

The proof of the above proposition is straightforward and is omitted. The economic point here is that whereas the correlation between dividends and return volatility, $\rho_{dx}$, represents the leverage effect, the difference $\rho_{rx}(x) - \rho_{dx}$ represents the volatility feedback effect. 

Also \citet{Ang07} provide a solution for the price-dividend ratio under a continuous stochastic volatility model and ICAPM (see Corollary 3.6 and Section 3.6 therein). Their solution, however, differs substantially from ours. First, we assume that {\em total} return volatility follows the Stein-Stein or Heston type process, whereas in \citep{Ang07}, squared total return volatility comprises (i) squared dividend growth volatility and (ii) a residual component, in which only the residual component follows the Heston model. The residual component represents, in fact, the volatility arising from time-varying discount rates. Second, and more importantly, in our model investors are rewarded for {\em total} return volatility as the original ICAPM predicts, whereas in \citep{Ang07}, investors are rewarded for residual volatility (discount rate volatility), but not for dividend growth volatility. Technically, their specification yields a closed-form solution, but economically such an assumption is questionable as the dividend growth risk remains unpriced. Interestingly, some papers in the volatility feedback literature assume that the risk premium depends only on dividend growth volatility instead of total return volatility, which is exactly the opposite to what Ang and Liu assume \citep*[see, e.g.,][]{Campbell92}. Third, their solution is based on a condition $f(0) = C$, which they refer to as the price-dividend ratio at time $t=0$ and match with the unconditional price-dividend ratio. However, since $f$ depends only on the state variable $x$ (also under their settings), $f(0)$ should refer to the price-dividend ratio with $x=0$ rather than with $t=0$, i.e., with zero volatility, not with zero time.\footnote{Remember that under the tranversality condition, $f$ is the function of volatility only, and calendar time is thus an irrelevant variable here.} Therefore, $f(0)$ cannot {  be} thought to represent the unconditional price-dividend ratio. Moreover, $f(0) = C$ cannot be assumed to be an exogenously determined constant and independent of $\gamma$ and other parameters. Indeed, as Figure \ref{FIG:f(x)} shows, $f(0)$ depends negatively on $\gamma$, which is very intuitive: return volatility is mean-reverting and never remains constantly at zero; therefore, the greater the price of diffusion return risk, the lower is the price-dividend ratio for a given instantaneous volatility because the future dividends are discounted at a higher rate. Note that under our model no expression is needed for $f(0)$, since we can use $f'(0) = 0$ as a boundary condition for the interval of $x \geq 0$, as imposed by the differentiability. Overall, Ang and Liu's characterization carries implications that are essentially different from ours.

\subsection{Why return volatility can be greater than dividend volatility}

At this point, let us consider the relation between dividend growth volatility and return volatility. The early literature offers much evidence that return volatility is greater than dividend growth volatility, i.e., $x^2 > \y(x)^2$. Suppose that the conditions of Proposition \ref{PROP:y} hold. Then it is easy to show that $x^2 > \y(x)^2$ if and only if
  \[
  \rho_{dx} < -\frac{\sigma_x}{2x}\frac{f_x(x)}{f(x)}.
  \]
  Because squared return volatility always has a non-positive effect on the price-dividend ratio, i.e., $f_x(x) \leq 0$ for all $x \geq 0$ and $f_x(x) \geq 0$ for all $x \leq 0$, the right hand side of the above inequality is non-negative. Consequently, if the correlation between dividends and return volatility is zero or less, return volatility is higher than dividend growth volatility. On the other hand, if the correlation between dividends and return volatility is positive and high enough, then return volatility can be lower than dividend growth volatility. 

\begin{figure}[h]
\begin{center}
  \includegraphics*[width=1\textwidth]{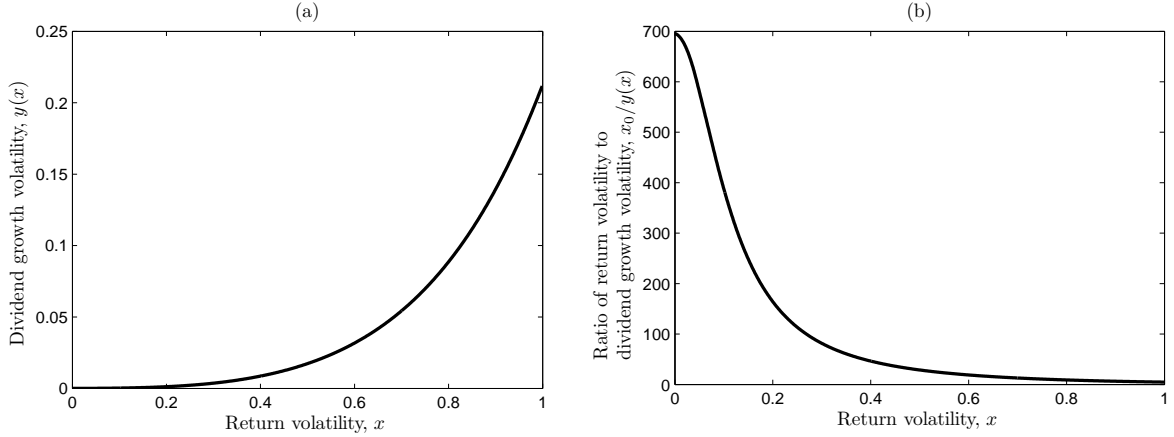}
  \caption{\linespread{1} \footnotesize{{  Can high return volatility plausibly be explained? The figure demonstrate a possible relation of dividend growth volatility, $\y (x)$, with respect to return volatility, $x$, under teh volatility feedback effect}. The parameters are $\gamma = 3.115$, $\alpha = 0.08$, $\sigma_x = 0.2$, $\beta = 0.5$, $r = 0.02$, and $\rho_{dx} = -0.5$. In (a), dividend growth volatility is plotted against return volatility, and in (b) the ratio of return volatility to dividend growth volatility is plotted.}}
\label{FIG:sigma_d(x)_extreme}
\end{center}
\end{figure}

According to our model, the ratio of squared return volatility to squared dividend growth volatility can be very large. If we calculate the ratio using the same parameter values as in Figure \ref{FIG:f(x)}, our model can yield an extremely high ratio of return volatility to dividend growth volatility, with $\gamma=3$ and $\alpha = 0.08$ {  the ratio is} even higher than 10 for all $0 \leq x < 0.5$. In fact, if $\gamma$ is high enough, the ratio can be infinite. In the light of Eq.\ (\ref{EQ:y(x)}), such a relation is easy to understand mathematically: for $\rho_{dx} \leq 0$, $\y(x)$ would approach zero for strictly positive $x$, if $(f_x/f)^2$ were high enough (due to a relatively high $\gamma$). In fact, $\y(x)$ could even be non-negative for a strictly positive $x$, but this would violate Assumption \ref{ASS:dD} ({\em ii}). Economically, this means that volatility feedback almost alone explains the return variance
  \[
  x^2 = \y(x)^2 + \sigma_{x}^2 \frac{f_x(x)^2}{f(x)^2} + 2 \rho_{dx} \y(x) \sigma_x \frac{f_x(x)}{f(x)}.
  \]
In other words, if $\gamma$ is relatively high, volatility feedback can amplify a very small but nonzero dividend growth volatility to a relatively high return volatility. This extreme situation is illustrated in Figure \ref{FIG:sigma_d(x)_extreme} by increasing $\gamma$ to $3.115$. According to plot (b), with these parameter values the ratio of return volatility to dividend growth volatility can be as high as 700. 

Recently, \citet{Bali10} estimated the risk aversion coefficient to be even more than {  4 and to be} highly significant for S\&P100 stocks \citep*[see also][]{Bali08}. In the light of these empirical estimates of $\gamma$, we argue that the high ratio of return volatility to dividend growth volatility, or the ``excess'' volatility of the stock market, can be explained {  by} volatility feedback. Our finding is in sharp contrast to the conventional argument that stock market volatility is "too" high \citep*[i.e. equity volatility puzzle, see, e.g.][]{Shiller81}. Shiller, however, assumed that the cost of capital is deterministic and unaffected by return volatility. If the discount rate varies, the stock price can vary even with unchanged dividends, and hence one should adopt time-varying discount rates. This assumption of a constant discount rate was later relaxed, e.g., by \citet{Cochrane92}. By modeling the stochastic cost of capital, we assert that the ratio of return volatility to dividend growth volatility can be arbitrarily large.

This study also shows that using {  the} log-linear approximation does not necessarily make sense under our settings. In particular, we could approximate the price-dividend ratio as
\[
\tilde{f}(x) = C_1 \exp\left(-C_2 x^2\right),
\]
where $C_1, C_2 > 0$ are constants. A similar approximation is applied, e.g., by \citet{Campbell02} and \citet{Bollerslev11b}.  Note that {  if} $\tilde{f}_x/\tilde{f} = -2C_2 x$, and the implied approximation of dividend growth volatility is linear in return volatility. However, the above figures show that this is not the case under our model.

\subsection{Price dynamics revised}
We have now shown that under the transversality condition the price-dividend-ratio depends on return volatility and is hence stochastic. Consequently, dividend yield cannot be assumed constant in time, as is typically done in the financial literature. In addition, the correlation between returns and return volatility is endogenously determined and stochastic, even though fairly stable. Instead of assuming a constant dividend yield and a constant correlation, the stock price process could be more appropriately written as follows:
\begin{equation}\begin{split}\label{EQ:dP_revised}
  \ud P_{t} &=  \left(r + \gamma x_t^2 - \frac{1}{f(x_t)} \right) P_t \ud t + x_t P_t \ud B_t^r\\
  &=  \left(r + \gamma x_t^2 - \frac{1}{f(x_t)} \right) P_t \ud t + \y(x_t) P_t \ud B_t^d + \sigma_x \frac{f_x(x_t)}{f(x_t)} P_t \ud B_t^x,
\end{split}\end{equation}
where $f(x)$ satisfies Eq.\  (\ref{EQ:DE}). Note that the random term, $x_t \ud B_t^r$, can be decomposed into a dividend term, $\y(x_t) \ud B_t^d$, and a volatility term, $ \sigma_x \frac{f_x(x_t)}{f(x_t)} \ud B_t^x$; i.e., the price process is driven by changes in dividends and return volatility. Moreover, return volatility was assumed to {  evolve} according to Equation (\ref{EQ:dx}). 

To understand how this expression captures volatility feedback, suppose that return volatility is positive,  $x_t > 0$, and that it increases, $\ud x_t > 0$. The first observation is that the expected rate of return, $r + \gamma x_t^2$, increases. The positive change in the drift term is, however, lessened or even reversed by a change in the dividend yield. As demonstrated earlier, an increase in squared return volatility decreases the price-dividend ratio, or, in other words, increases the dividend yield, $1/f$, and thus potentially decreases the expected price appreciation, resulting in a pull-down. In addition, under these conditions, the last term on the right hand side in Eq.\ (\ref{EQ:dP_revised}, second line) is strictly negative, further decreasing the stock price. Therefore, an increase in squared volatility results in three effects on the stock price: an increased expected rate of return, an increased dividend yield, and a negative random shock. Moreover, because dividend growth volatility increases together with return volatility, the stock price becomes more sensitive to dividend shocks.

The volatility process follows the mean-reversion Ornstein-Uhlenbeck process, which provides an exact solution that we can simulate with arbitrary time steps; however, simulation of the stock price process is appropriate only with short time {  steps}:
\[\begin{split}
  P_{t + \Delta t} &=  P_t \exp\left[ \left(r + \gamma x_t^2 - \frac{1}{f(x_t)} - \frac{1}{2}x_t^2 \right) \Delta t + \y(x_t) \sqrt{\Delta t} \epsilon_t^d + \sigma_x \frac{f_x(x_t)}{f(x_t)} \sqrt{\Delta t} \epsilon_t^x \right],\\
x_{t + \Delta t} &= x_t \exp \left(-\beta \Delta t \right) + \sigma_x \sqrt{\frac{1-\exp\left(-2 \beta \Delta t\right)}{2 \beta}} \epsilon_t^x,
\end{split}\] 
where $\epsilon^d, \epsilon^x \sim N(0,1)$, $\mathrm{Corr}(\epsilon^d, \epsilon^x) = \rho_{dx}$. In each step, $f(x)$ and $f_x(x)$ can be solved for a given $x$ with Proposition (\ref{PROP:DE}) and $\y(x)$ with Proposition \ref{PROP:y}. Note that the dividend process can be determined from $D_t = P_t/f(x_t)$ or, alternatively, simulated directly:
\begin{equation}\label{EQ:D_simulation}
D_{t + \Delta t} =  D_t \exp\left[ \left(\alpha - \frac{1}{2}\y(x_t)^2 \right) \Delta t + \y(x_t) \sqrt{\Delta t} \epsilon_t^d \right].
\end{equation}
When the same random number sequences are used and the time steps shrink, both approaches yield identical dividend stream sequences with $D_0 = P_0/f(x_0)$.  In addition, instead of simulating the return process directly, the dividend stream and return volatility can also be simulated together with the  stock price, which is then determined using the relation $P_t = D_t f(x_t)$. 

\begin{figure}[h]
\begin{center}
\includegraphics*[scale=0.77]{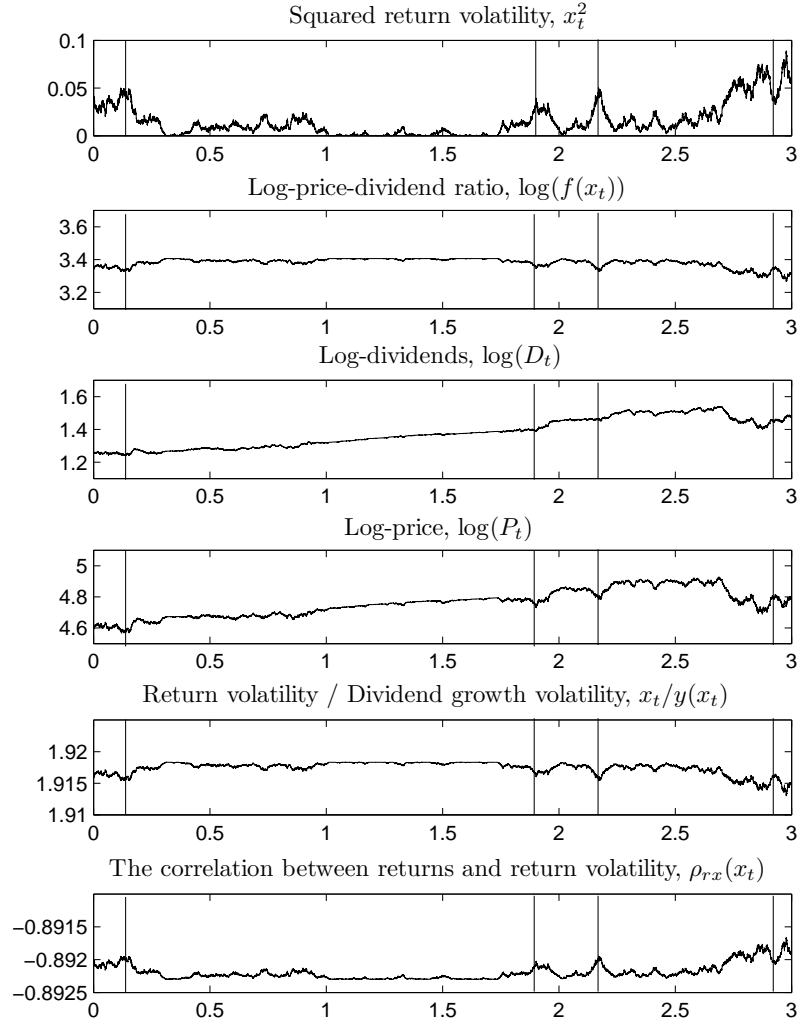}
\caption{\linespread{1} \footnotesize{Sample paths. The parameters are $\sigma_x = 0.2$, $\beta = 0.5$, $r = 0.02$, $\alpha = 0.05$, and $\rho_{dx} = -0.5$. Moreover, $x_0 = 0.2$, $P_0 = \$100$, and $\Delta t = 1/(24 \times 252)$ years.}}
\label{FIG:Paths}
\end{center}
\end{figure}

Sample paths of squared return volatility, log-price-dividend ratio, log-dividends, and log-price are illustrated in Figure \ref{FIG:Paths}. The prices, dividends, and their ratios are given in logarithmic form with equally stepped tics because, by the definition of price-dividend ratio, $\ln(P) = \ln(D) + \ln(f)$. The figure also plots the ratio {  of} return volatility to dividend growth volatility and the correlation between returns and return volatility. We have highlighted four time instants with large movements in return volatility with stable dividends. Clearly, the price-dividend ratio, and then also the stock price, has reacted negatively ({  resp.} positively) to the positive ({  resp.} negative) volatility movements because of volatility feedback. For example, observe  the period before the third highlighted time instant, in which squared volatility increased to about 0.05 ($t \approx 2.17$). In this period, the dividend increased slightly and had practically no effect on stock price movements, yet the stock price fell substantially in response to an increase in the cost of capital. The fourth highlighted time-period ($t \approx 2.92$) represents a situation in which the stock price level increased due to {{  a} decrease in the cost of capital. When the price-dividend ratio (or equivalently, constant dividend yield) is assumed constant, as done traditionally, the stock price goes up if and only if dividends increase. However, our characterization {  allows} stock prices and dividends to move in opposite {  directions}. Moreover, our model implies that call prices and the underlying stock do not necessary move in the same direction, {  in agreement with what was} empirically observed by \citet{Bakshi00b}. In particular, a change in the call price can be positive or negative depending on which effect, increased squared volatility or increased dividend yield, predominates. Option pricing under our settings is discussed further in Section \ref{SEC:Option_Pricing}.

The figure illustrates that the modeled return volatility is approximately 1.915 times higher than {  the} dividend growth volatility, and that this ratio varies slightly. The estimated unconditional ratio, 1.907, is close to these theoretical values. In addition, the correlation between returns and return volatility is quite stable in time and {  flucturates} around -0.892. The estimated unconditional correlation between $\Delta(x^2)$ and $\Delta \ln P$ is -0.8868, again close to the theoretical values, and the estimated correlation between $\Delta(x^2)$ and $\Delta \ln D$ is -0.4967, which is very close to $\rho_{dx} = -0.5$. The difference between $\mathrm{Corr}(\Delta (x^2), \Delta \ln P)$ and $\mathrm{Corr}(\Delta (x^2), \Delta \ln D)$ can be justified with volatility feedback; in fact, the difference could be used as a measure of volatility feedback. Finally, the figure also demonstrates volatility clustering: periods of high ({  resp.} low) volatility are followed by high ({  resp.} low) volatility. Because return volatility and dividend growth volatility evolve hand in hand, volatility clustering is not only about returns but also about dividends.

\section{Option valuation}\label{SEC:Option_Pricing}

\subsection{Risk-neutral dynamics}
Under the risk-neutral probability measure $\mathbb{Q}$, cumulative stock return and return volatility {  evolve} as
\begin{align}
  \ud R_t &= r \ud t + x_t \ud \tilde{B}_t^r,\nonumber\\
  \ud x_t &= -\tilde{\beta}(x_t) x_t \ud t + \sigma_x \ud \tilde{B}_t^x, \label{EQ:dx_risk_neutral_diffusion}
\end{align}
where $\tilde{B}_t^r$ and $\tilde{B}_t^x$ are Brownian motions under the probability measure $\mathbb{Q}$, and $\tilde{\beta}$ is the speed of the mean reversion under $\mathbb{Q}$. We realize immediately that the above is satisfied if
\[\begin{split}
  \ud \tilde{B}_t^r &= \ud B_t^r + \gamma x_t \ud t,\\
  \ud \tilde{B}_t^x &= \ud B_t^x + \frac{\lambda_x(x_t)}{\sigma_x} x_t \ud t
\end{split}\]
with
\[
x_t \ud \tilde{B}_t^r = \y(x_t) \ud \tilde{B}_t^d + \sigma_x \frac{f_x(x_t)}{f(x_t)} \ud \tilde{B}_t^x,
\]
where $\lambda_x(x_t) = \tilde{\beta}(x_t) - \beta$ represents the volatility risk premium, which is non-zero for equity \citep*[see, e.g.,][]{Lamoureux93,Jiang99,Bakshi03}. For simplicity, we suppose that $\lambda_x = \tilde{\beta} - \beta $ is constant. To express the dividend process under the risk-neutral measure, we write
\[\begin{split}
  \y(x_t) \ud B_t^d &= x_t \ud {B}_t^r - \sigma_x \frac{f_x(x_t)}{f(x_t)} \ud {B}_t^x\\
  &= \y(x_t) \ud \tilde{B}_t^d - \left(\gamma x_t^2 -\frac{f_x(x_t)}{f(x_t)} \lambda_x x_t \right) \ud t,
\end{split}\]
which implies that the dividend process under the risk-neutral measure is given by
\begin{equation}\label{EQ:dD_risk_neutral_diffusion}
  \ud D_t = \left(\alpha - \gamma x_t^2 + \frac{f_x(x_t)}{f(x_t)} \lambda_x x_t \right) D_t \ud t + \y(x_t) D_t \ud \tilde{B}_t^d.
\end{equation}
Therefore, the rate of expected dividend growth becomes stochastic under the risk neutral measure. Note that the greater the $\gamma$, the less the expected dividend growth under the risk-neutral probability measure. Moreover, a negative price of volatility risk, $\lambda_x < 0$, affects the expected dividend growth positively under the risk-neutral measure.

It is also worth observing that the price-dividend ratio satisfies the same relation under both physical and risk-neutral probability measures. To see this, suppose that the conditions of Proposition \ref{PROP:y} hold. Then the stock price evolves under risk-neutral dynamics as follows:
\begin{equation}\label{EQ:dR_risk_neutral}\begin{split}
    \ud P_t &= \bigg( \alpha - \gamma x_t^2 + \lambda_x x_t \frac{f_x(x_t)}{f(x_t)}  + \left(
    \rho_{dx} \sigma_x \y(x_t) - \tilde{\beta} x_t \right) \frac{f_x (x_t)}{f(x_t)} + \frac{1}{2} \sigma_x^2 \frac{f_{xx}(x_t)}{f(x_t)} \bigg) P_t \ud t\\
    &\quad + \sigma_x \frac{f_x(x_t)}{f(x_t)} P_t \ud \tilde{B}_t^x +\y(x_t) P_t \ud \tilde{B}_t^d,
  \end{split}\end{equation}
where $\tilde{\beta} = \beta + \lambda_x$, and thus the terms including $\lambda_x$ cancel each other out. Under the risk-neutral measure, the expected rate of return of any asset equals the instantaneous risk-free interest rate. Now when we apply this principle and match the drift term of the risk-neutral price process (\ref{EQ:dR_risk_neutral}) with the expected price appreciation under the risk-neutral measure, $r - 1/f(x)$, Eq.\ (\ref{EQ:DE}) follows. This means that the stock is priced equivalently under both physical and risk-neutral probability measures.
Consequently, under $\mathbb{Q}$, the stock price follows 
\begin{equation}\label{EQ:dP_risk_neutral_diffusion}
  \ud P_t = \left(r - \frac{1}{f(x_t)} \right)P_t \ud t + \y(x_t) P_t \ud \tilde{B}_t^d + \sigma_x \frac{f_x(x_t)}{f(x_t)} P_t \ud \tilde{B}_t^x,
\end{equation}
where $f(x)$ satisfies Eq.\  (\ref{EQ:DE}) and can be solved for given $x$ with $r$ and $\alpha$ and structural parameters $\sigma_x, \beta, \gamma, \rho_{dx}$.

\subsection{Option prices}

Given the risk-neutral dynamics in (\ref{EQ:dP_risk_neutral_diffusion}) and (\ref{EQ:dx_risk_neutral_diffusion}), the price of a European call option can be computed as
\begin{equation}\label{EQ:C}
c(t, P_t, x_t, T, K, r, \alpha; \theta) = \exp(-r(T-t)) \mathbb{E}^Q_t \left[(P_T - K)^+ \right],
\end{equation}
where $T$ is the time of maturity, $K$ the exercise price, and $\theta = \{\sigma_x, \beta, \tilde{\beta}, \gamma, \rho_{dx} \}$ the set of structural parameters. Like, e.g., most GARCH models and the so-called VAR volatility model, our model requires Monte Carlo simulations to compute option prices \citep*[see, e.g.,][]{Christoffersen04,Christoffersen11,Barone08}. To speed up computations, we use antithetic variates and distributed computing. 

What is very fundamental here is that to compute the expected payoff in Eq.\ (\ref{EQ:C}), we {\em need} to know all the parameters $\theta = \{\sigma_x, \beta, \tilde{\beta}, \gamma, \rho_{dx} \}$. Specifically, to compute the right hand side of Eq.\ (\ref{EQ:C}) with Monte Carlo methods, we simulate the discretized risk-neutral processes
\[\begin{split}
P_{t + \Delta t} &=  P_t \exp\left[ \left(r - \frac{1}{f(x_t)} - \frac{1}{2}x_t^2 \right) \Delta t + \y(x_t) \sqrt{\Delta t} \tilde{\epsilon}_t^d + \sigma_x \frac{f_x(x_t)}{f(x_t)} \sqrt{\Delta t} \tilde{\epsilon}_t^x \right],\\
x_{t + \Delta t} &= x_t \exp \left(-\tilde{\beta} \Delta t \right) + \sigma_x \sqrt{\frac{1-\exp\left(-2 \tilde{\beta} \Delta t\right)}{2 \tilde{\beta}}} \tilde{\epsilon}_t^x,
\end{split}\]
where $\tilde{\epsilon}^d, \tilde{\epsilon}^x \sim N(0,1)$, $\mathrm{Corr}\left(\tilde{\epsilon}^d, \tilde{\epsilon}^x \right) = \rho_{dx}$. Even though parameters $\gamma, \beta$, and $\alpha$ do not directly appear in the above expressions, they are necessary, together with $r, \sigma_x$, and $\rho_{dx}$, to compute $f(x)$ in each time step for a given $x$. Thus {\em future stock price distributions and {  hence} also option prices depend on the price of diffusion return risk,} $\gamma$. This is in very sharp contrast to the derivative pricing literature, which, following \citet{Black73}, considers the price of diffusion return risk and the stock's expected rate of return irrelevant to option pricing. However, in our dynamics, $\gamma$ affects option prices, mainly via dividend yield, $1/f(x)$, which is expressed as a function of return volatility with $\gamma$ as a structural parameter of that function. In other words, $\gamma$ determines the sensitivity of dividend yield to return volatility and thereby affects option prices. In some current option pricing models, option prices can be seen as dependent on the price of the return risk, but these models differ essentially from ours. To understand the difference, consider, for example, the \cite{Heston00} model, in which $\theta$ is the leverage parameter under the physical measure whereas under the risk-neutral measure it is $\tilde{\theta} \equiv \theta + \gamma$, where $\gamma$ is the price of the return risk. Hence, one could say that for fixed $\theta$, a change in $\gamma$ affects option prices through $\tilde{\theta}$. However, to price options, all we need is the combination $\theta + \gamma$; consequently, we cannot separately identify $\theta$ and $\gamma$ when estimating the Heston-Nandi model using option data alone under the risk-neutral probability measure, because we can only estimate the combination $\tilde{\theta} \equiv \theta + \gamma$ \citep*[see the similar discussion {  of} the Leverage model][on-line Appendix]{Christoffersen04}. Moreover, because options are priced in terms of the risk-neutralized volatility process, under the standard Heston model (with two parameters) option prices directly depend on the sum of $\tilde{\beta} = \beta + \lambda_x$ and only the combination of $\beta + \lambda$ can be estimated with historical return data. However, under our dynamics we need both parameters, $\beta$ and $\tilde{\beta}$, not only their sum, to price options. That is, we need {\em separately} both physical and risk-neutral parameters, or, in other words, we need the market price of the diffusion return risk and volatility risk-premium; this arrangement allows us to estimate the forward-looking, option-implied market price of the return risk and the volatility risk-premium using option data alone. 

Figure \ref{FIG:c_gamma_wrt_T_at-the-money} illustrates the relation between option prices and the price of diffusion return risk. {  The parameter values are expressed in the figure caption. In plot (b), we perform a sensitivity analysis by varying $\beta$ and $\sigma_x$ from Bakshi's estimates \cite{Bakshi97}.\footnote{Note that the values are the same as used in Figure \ref{FIG:f(x)}b, except that $\alpha = 0.015$. This is justifiable because for $\gamma = 0$, the stock price is well defined only if $r > \alpha$.}} The option prices were calculated with Monte Carlo simulations using antithetic variates and 10,000 paths/option. 
Clearly, the greater the price of diffusion return risk, the greater the dividend yield and the lower the call price. Moreover, the greater the time to maturity, the greater the effect of an increase in the price of diffusion return risk on the call price. {  In addition, option prices are higher with $\rho_{dx} = -0.5$ than with $\rho_{dx} = 0.5$, and correspondingly, initial dividend yield is lower with with $\rho_{dx} = -0.5$ than with $\rho_{dx} = 0.5$. A doubling value of $\beta$ and a corresponding change in $\sigma_x$ do not affect the results substantially, and the shape of the curves remain the same.}

\begin{figure}[h]
\begin{center}
  \includegraphics*[width=1\textwidth, angle=0]{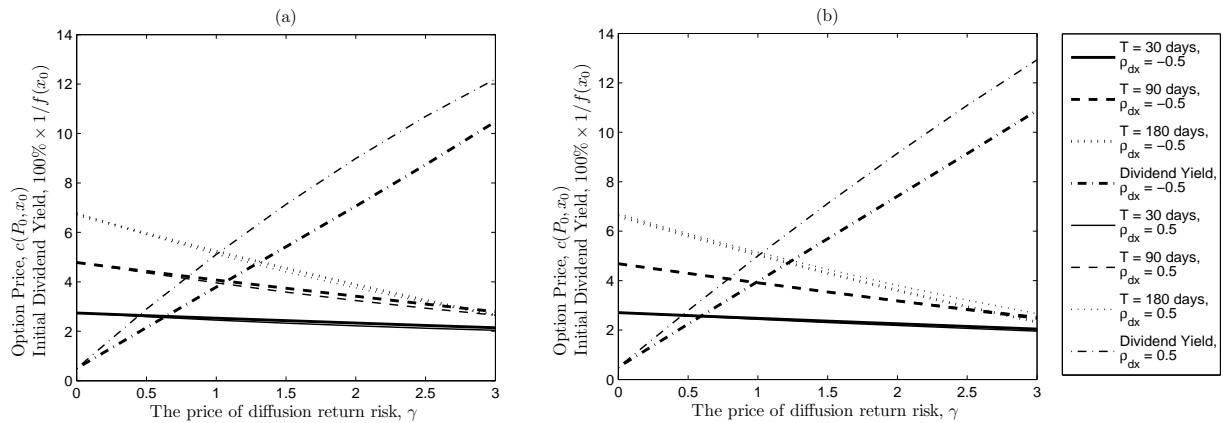}
  \caption{\linespread{1} \footnotesize{  Effect of the price of diffusion return risk on European call option prices. Parameters are $r = 0.02$, $\alpha = 0.015$, and $\rho_{dx} = -0.5$ (thicker lines) or $\rho_{dx} = 0.5$ (thinner lines). In plot (a), $\beta = \tilde{\beta} = 0.5$, $\sigma_x = \sqrt{2\theta\beta} = 0.2$ with $\theta = 0.04$, and in plot (b), $\beta = \tilde{\beta} = 1$, $\sigma_x = \sqrt{2\theta\beta} \approx 0.2828$ with $\theta = 0.04$. Moreover, $x_0 = 0.2$, $P_0 = \$100$, $K=\$100$, and $\Delta t = 1/252$ year.}}
\label{FIG:c_gamma_wrt_T_at-the-money}
\end{center}
\end{figure}

Figure \ref{FIG:c_lambda_wrt_T_at-the-money} shows how option prices depend on the volatility risk premium. In plot (a), we fix $\tilde{\beta}$ and vary $\beta = \tilde{\beta} - \lambda_x$ whereas in (b) $\beta$ is fixed and $\tilde{\beta} = \beta + \lambda_x$ varies. Plot (a) shows also the dividend yield whereas in (b) it is constant ($\tilde{\beta}$ does not appear in Eq.\ \ref{EQ:DE}) and thus ignored. In both cases, an increase in $\lambda_x$, which represents the difference of $\tilde{\beta}$ and $\beta$, decreases option prices. {  Again, a doubling value of $\beta$ and a corresponding change in $\sigma_x$ do not affect the results substantially.}

\begin{figure}[h]
\begin{center}
  \includegraphics*[width=1\textwidth]{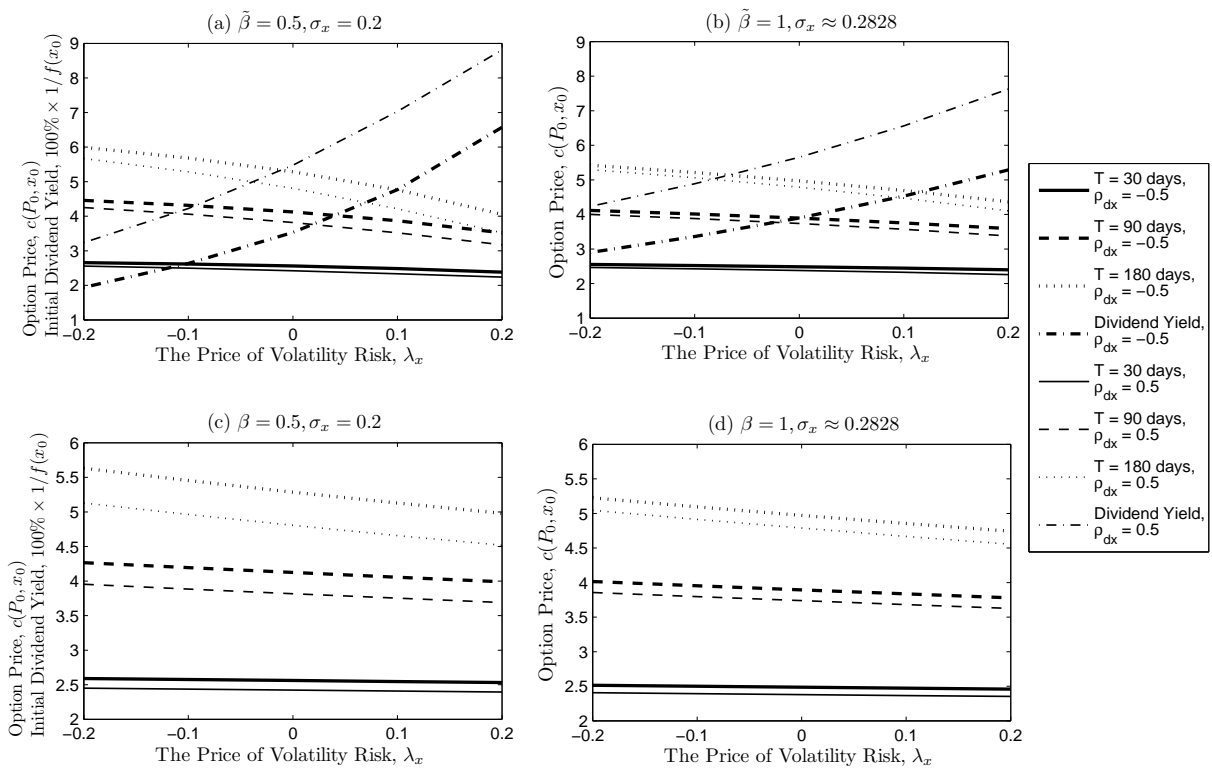}
  \caption{\linespread{1} \footnotesize{  Effect of the price of volatility risk on European call option prices. Parameters are $\gamma = 2, r = 0.02$, $\alpha = 0.05$, and $\rho_{dx} = -0.5$ (thicker lines) or $\rho_{dx} = 0.5$ (thinner lines). In plot(a) $\tilde{\beta}= 0.5$, $\sigma_x = 0.2$ and in (c), $\beta = 0.5$, $\sigma_x = 0.2$, and in plots (b) $\tilde{\beta} = 1$, $\sigma_x \approx 0.2828$ and (d), $\beta = 1$, $\sigma_x \approx 0.2828$. On the other hand, in (a) and (b), $\tilde{\beta}$ is fixed whereas in (c) and (d), $\beta$ is fixed (remember that $\lambda_x = \tilde{\beta} - \beta$). Moreover, $x_0 = 0.2$, $P_0 = \$100$, $K=\$100$, and $\Delta t = 1/252$ year.}}

\label{FIG:c_lambda_wrt_T_at-the-money}
\end{center}
\end{figure}

Because in our model options prices depend on  {\em all} the parameters $\{\sigma_x, \beta,$ $\tilde{\beta}, \gamma, \rho_{dx} \}$ in addition to the risk-free interest rates and the expected dividend growth rate, the parameters (including $\gamma$ and $\beta$) could be estimated using information on option prices by directly minimizing pricing errors. This also contradicts the early empirical literature on financial economics, where the coefficient of risk relative aversion, $\gamma$, and the speed of the ``physical'' volatility mean-reversion, $\beta$, have so far been estimated using time series of asset returns, not pure option prices.\footnote{Note that under the original Heston model (or other current stochastic volatility models), the price of diffusion return risk and the speed of the mean reversion of physical processes can be estimated only by using time series of asset returns, not option prices, since Heston's option prices are not affected by ``physical parameters.'' } Therefore, we are motivated to estimate the parameters of our model using information on options prices alone, and we will use this approach in the empirical section of this study. 

At this point, it is worth pointing out that the original Heston model (with the Ornstein-Uhlenbeck process) can be seen as a special case of ours. For $\gamma = 0$, we get that $f = 1/(r - \alpha)$ is constant with respect to $x$, implying a constant dividend yield $r - \alpha$ and equal dividend and return volatilities, $\y(x) = x$. By denoting $B_t^d$ by $B_t$ and $\rho_{dx}$ by $\rho$, the risk-neutral versions of the stock price and the volatility process with $\gamma=0$ can be expressed as
\[\begin{split}
  \ud P_t &= \alpha P_t \ud t + x_t P_t \ud \tilde{B}_t,\\
  \ud x_t &= -\tilde{\beta} x_t \ud t + \sigma_x \ud  \tilde{B}_t^x,
\end{split}\]
where $\ud B_t \ud B_t^x = \rho \ud t$, which corresponds exactly to Heston's original specification. Therefore, our model should be able to price options at least as well as Heston's model with the dividend yield of $r - \alpha$, or even better, as we will see in the empirical part of this study. In fact, we could say that under Merton's ICAPM, Heston's characterization of a constant dividend yield implicitly assumes $\gamma = 0$, viz. risk-neutral investors. Note also that if we set $\gamma = 0$, option prices no longer depend on the speed of the physical mean revision, $\beta$, which cannot be estimated using pure option data.

One interesting implication of our characterization is that the option price can be a decreasing function of squared return volatility. The reason here is that an increase in squared volatility increases the dividend yield, and thus potentially lowers the call price. The early literature usually argues the opposite, i.e., that because of the convex payoff, return volatility has a positive effect on the standard call option \citep*[see, e.g.,][]{Merton73b,Bergman96,Hobson98,Janson02,Kijima02}. These arguments for a positive relation between call price and volatility are, however, based on the assumption of a constant dividend yield or absence of dividends.

To see how return volatility affects options in our model, let us, for simplicity, first consider a special case of $K \downarrow 0$; i.e., the exercise price is zero and the 'option' holder gets the underlying stock for free at time $T$. In this case, the option price is
\begin{equation}\begin{split}\label{EQ:C_K=0}
  c(t, P_t, x_t, T, 0, r, \alpha; \theta) &= \exp(-r(T-t)) \mathbb{E}^Q_t P_T\\
    &= P_t \mathbb{E}^Q_t \exp \left[\int_t^T \left( - \frac{1}{f(x_s)} - \frac{1}{2}x_s^2\right) ds + \int_t^T x_s \ud \tilde{B}_s^r  \right].
\end{split}\end{equation}
The price of this option is less than that of the underlying stock, because the option holder receives no dividends until maturity, and hence the underlying price is reduced by the expected cumulative dividends. If the current squared return volatility $x_t^2$ increases, then not only the current dividend yield but also the expected dividend yields increase because of the persistence of stock return volatility; consequently, the expected terminal price, $\mathbb{E}_t^Q P_T$, is lower. In fact, we  suppose here that the stock price level is not affected by an increase in volatility, which, according to Definition \ref{DEF:f},  must mean that the current level of dividends must increase to respond to a lower price-dividend ratio. Greater dividends mean a greater shortfall for the option holder, reducing the option price. This is illustrated in figure \ref{FIG:c_x_K=0}, plot (a).  The same could also be put differently. We could think of a situation in which the stock price level reacts to an increase in squared return volatility, while the level of dividends remains an exogenous variable and unaffected, just as the theory of volatility feedback predicts. We can then express the option contract in the terms of dividends, rather than a function of the spot price:
\[\begin{split}
  c(t, f(x_t) D_t, x_t, T, 0, r, \alpha; \theta) &= f(x_t) D_t \mathbb{E}^Q_t \exp \left[\int_t^T \left( - \frac{1}{f(x_s)} - \frac{1}{2}x_s^2\right) ds + \int_t^T x_s \ud \tilde{B}_s^r  \right].
\end{split}\]
Because $f(x)$ is decreasing w.r.t. $x$, an increase in squared return volatility has two negative effects on option prices: via increased dividend yield and via a lower stock price level. As figure \ref{FIG:c_x_K=0}, plot (b) shows, squared return volatility can then substantially decrease the option price. {  In addition, the grater the speed of adjustment of the volatility process ($\beta$), the less is the effect of initial volatility on option prices as squared volatility is pushed toward the average volatility level faster.}

\begin{figure}[h]
\begin{center}
  \includegraphics*[width=1\textwidth]{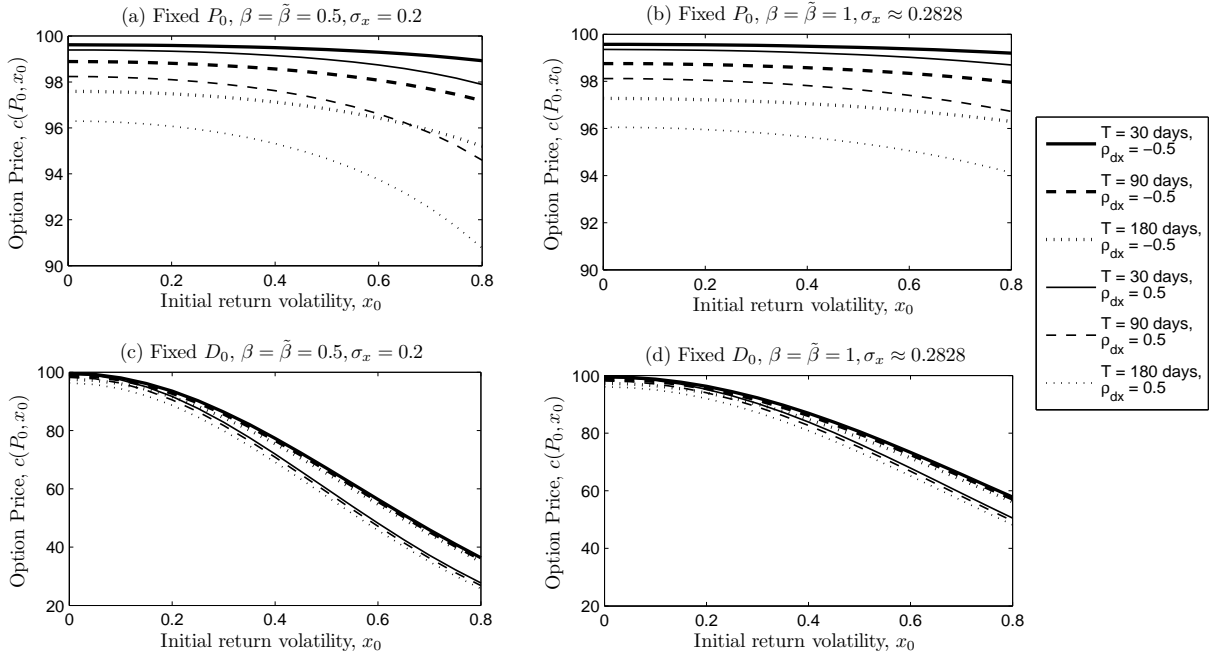}
  \caption{\linespread{1} \footnotesize{  Effect of return volatility on the price of a European call option with $K=\$0$. Parameters are $\gamma = 2, r = 0.02$, $\alpha = 0.05$, and $\rho_{dx} = -0.5$ (thicker lines) or $\rho_{dx} = 0.5$ (thinner lines). In plots (a) and (c), $\beta = \tilde{\beta} = 0.5$, $\sigma_x = \sqrt{2\theta\beta} = 0.2$ with $\theta = 0.04$, and in plots (b) and (d), $\beta = \tilde{\beta} = 1$, $\sigma_x = \sqrt{2\theta\beta} \approx 0.2828$ with $\theta = 0.04$. On the other hand, in (a) and (b), $P_0 = \$100$ is fixed whereas in (c) and (d), $D_0 = \$100/f(0) \approx \$3.3165$ is fixed. Moreover, $x_0 = 0.2$ and $\Delta t = 1/252$ year.}}
\label{FIG:c_x_K=0}
\end{center}
\end{figure}

With strictly positive exercise prices, $K>0$, the payoff becomes convex, and thus an increase in squared volatility can also increase the option price, depending on which effects, positive or negative, dominate. Intuitively, the greater the $K$, the greater the ``convexity effect'' and the more squared return volatility can increase option prices. Figure \ref{FIG:c_x_at-the-money} illustrates how the price of an option with $K = \$100$ changes with return volatility. In plot (a), we keep the current spot price fixed whereas in (b) current dividends are fixed. In the first case, convexity dominates the increased dividend yield, whereas in the latter case, the option price can be a non-monotonic function of return volatility. Note that in (a) the option is at-the-money for all $x \geq 0$, whereas in (b) it is at-the-money only for $x=0$ and out-of-the-money for all $x>0$. {  Changes in $\beta$ and $\sigma_x$ have a negligible effect on the relation between return volatility and option prices.}

\begin{figure}[h]
\begin{center}
  \includegraphics*[width=1\textwidth]{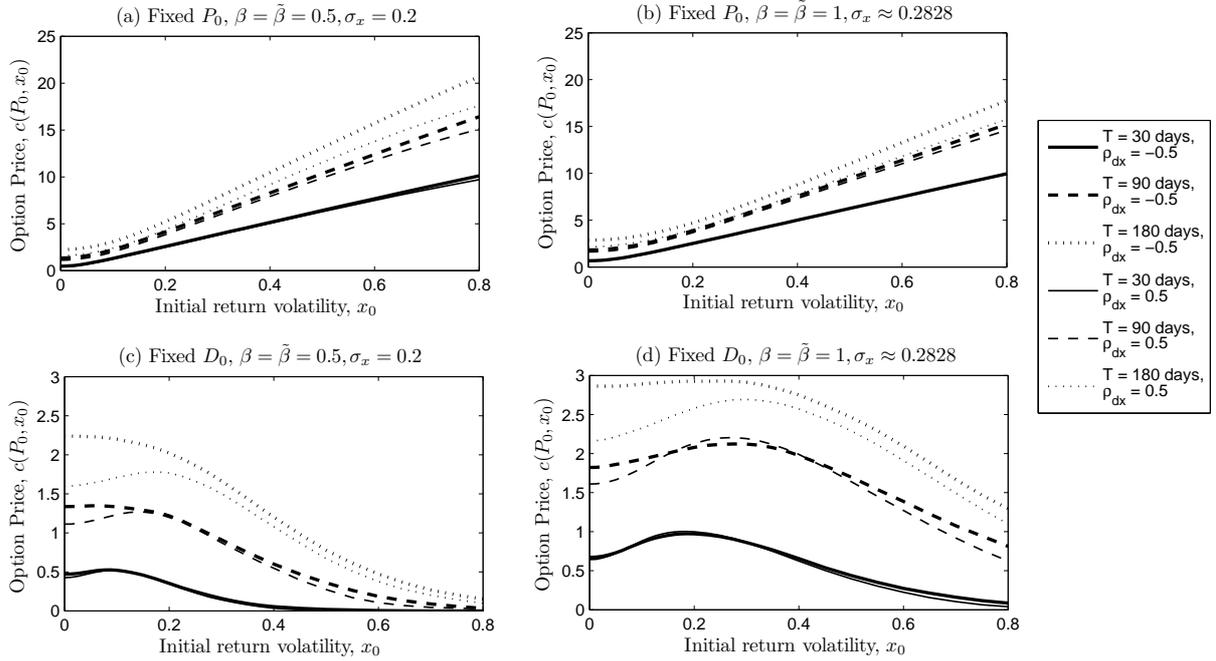}
  \caption{\linespread{1} \footnotesize{  Effect of return volatility on the price of a European call option with $K=\$100$. Parameters are $\gamma = 2, r = 0.02$, $\alpha = 0.05$, and $\rho_{dx} = -0.5$ (thicker lines) or $\rho_{dx} = 0.5$ (thinner lines). In plots (a) and (c), $\beta = \tilde{\beta} = 0.5$, $\sigma_x = \sqrt{2\theta\beta} = 0.2$ with $\theta = 0.04$, and in plots (b) and (d), $\beta = \tilde{\beta} = 1$, $\sigma_x = \sqrt{2\theta\beta} \approx 0.2828$ with $\theta = 0.04$. On the other hand, in (a) and (b), $P_0 = \$100$ is fixed whereas in (c) and (d), $D_0 = \$100/f(0) \approx \$3.3165$ is fixed. Moreover, $x_0 = 0.2$ and $\Delta t = 1/252$ year.}}
\label{FIG:c_x_at-the-money}
\end{center}
\end{figure}

\section{Calibration to option prices}\label{SEC:Empirical_Analysis}

\subsection{Data and methodology}

In this section, by calibrating our model to the sample of S\&P500 call options, we illustrate how it is possible to obtain forward-looking estimates for the price of diffusion return risk using information on option contracts alone without the need of the series of asset returns. At the same time, this paper is among the first attempts to obtain the option-implied values for the price of diffusion return risk, $\gamma$, and the volatility risk premium, $\lambda_x$. The use of information on {  option } contracts (instead of time series data) is motivated also by \citet{Christoffersen04} who argue, ``for the purpose of option valuation, it may be preferable to estimate the parameters directly using \dots option prices.'' \citet{Bakshi97}, among others, employ this estimation methodology using loss functions to minimize the pricing error of options. Since their work, a wealth of literature has appeared on evaluation of stochastic volatility models using empirical information on option prices \citep*[see, e.g.,][]{Chernov00,Christoffersen04,Huang04,Barone08,Christoffersen11b}. 

In this study, we use the sample of the daily data of S\&P 500 index call options traded on the Chicago Board Options Exchange (CBOE), and in particular the mid-point bid-ask quotes. Practitioners often estimate the coefficients of the underlying dynamics on observed option prices through static daily calibration, but instead of using single-day data, we use multi-day data. The option prices are samples of every trading day in 1995, from January 3 through December 27, a total of 21,166 observations.\footnote{The data were graciously provided by Gurdip Bakshi.} Data from 1995, and the 1990s data in general, is widely used in the literature \citep*[e.g.,][]{Bakshi03,Christoffersen04,Christoffersen08b,Bakshi10b,Christoffersen11b}, and our empirical results are thus comparable with those of recent studies on option pricing. The recorded S\&P 500 index values are not closing values but rather from the moment an option bid-ask quote was recorded. We use the data on daily three-month Treasury bill discounts and convert them first to annualized continuously compounded interest rates to price options each trading day. Moreover, we estimate the average growth rate of dividends, $\bar{\alpha} \approx 6.13\%$, from monthly data on S\&P 500 dividends\footnote{ The monthly dividend data were made available by Professor Robert Shiller on his web site \url{http://www.econ.yale.edu/~shiller/data.htm}.} from January 1995 through December 1996, covering the lengths of all the option contracts. A typical approach to take dividends into account is to calculate their present values until the maturity of each option and subtract them from the spot prices \citep*[see, e.g.,][]{Bakshi97,Christoffersen04,Christoffersen08b}. However, in our paper, dividend yield is determined endogenously by stock return volatility. 

We used some exclusionary criteria to filter out option data that could complicate calibration. From the original data set we excluded price data with a time stamp later than 3 p.m.; excluded options with maturity less than 6 days; excluded market prices below 3/8 dollars; and excluded options in conflict with the no-arbitrage rules. These criteria were also used by \citet{Bakshi97}, and similar criteria were used by \citet{Heston00}. 
To satisfy the arbitrage restriction, the option price must fall between the upper and lower bounds. To be precise, we must ensure that both, ask and bid quote, are between those bounds. Therefore, we test the ask quotes against the upper bound rule and the bid price against the lower bound rule. The lower bound of the no-arbitrage rule is $C_t \geq P_t - \mathrm{PVDIV} - Ke^{-r(T-t)}$, where PVDIV is the present value of dividends during the life of the option. Here we used dividend data in a traditional way to calculate the lower bounds of option prices. The upper bound is simply $C_t \leq P_t$.

After applying all the above criteria, our test data set contained 18,587 quotes. We divided the data set into two samples. First, the period from June 3, 1995 to August 4, 1995, covering the first 150 trading days and denoted as Sample A, was used exclusively for in-sample estimation. Second, we used the period between August 7, 1995 and December 29, 1995, a total of 100 trading days, as an out-of-sample data set, referred to as Sample B. Table \ref{TAB:Sample_A&B} shows the properties of the option data in detail.

\begin{table}[h!] 
\footnotesize{
\centering
\subtable[In-Sample Data]{
\centering
\resizebox{7.5cm}{!}{
\begin{tabular}{ccccc}
\toprule
            Moneyness & \multicolumn{ 3}{r}{Maturity (days to expiration)} &            \\
\midrule
           P/K &       $< 60$ &   60 - 180 & $\geq 180$ &   Subtotal \\
\midrule
$<0.94$ &        104 &        757 &        327 &       1188 \\
\       &   $(0.73)$ &   $(2.98)$ &   $(6.86)$ &            \\
0.94-0.97 &        710 &        690 &        115 &       1515 \\
\       &   $(1.68)$ &   $(7.81)$ &  $(18.45)$ &            \\
0.97-1.00 &       1196 &        780 &        117 &       2093 \\
\       &   $(4.94)$ &  $(14.86)$ &   $(27.1)$ &            \\
1.00-1.03 &       1170 &        744 &        213 &       2127 \\
\       &   $(13.75)$ &  $(24.69)$ &  $(35.74)$ &            \\
1.03-1.06 &       1075 &        732 &        173 &       1980 \\
\       &   $(25.86)$ &  $(35.97)$ &  $(44.83)$ &            \\
1.06 &       1473 &        815 &         76 &       2364 \\
\       &   $(43.56)$ &  $(49.53)$ &  $(55.75)$ &            \\
\midrule
        Subtotal &       5728 &       4518 &       1021 &       11267 \\
\bottomrule
\end{tabular}}
\label{tab: insampledata}
}
\subtable[Out-of-Sample Data ]{
\centering
\resizebox{7.5cm}{!}{
\begin{tabular}{ccccc}
\toprule
Moneyness & \multicolumn{ 3}{r}{Maturity (days to expiration)} &            \\
 \midrule
P/K &       $< 60$ &   60 - 180 & $\geq 180$ &   Subtotal \\
 \midrule
$<0.94$ &         53 &        139 &         66 &        258 \\
\       &   $(0.7)$ &   $(3.49)$ &  $(10.37)$ &    \\
0.94-0.97 &        195 &         92 &         35 &        322 \\
\       &   $(1.97)$ &   $(7.31)$ &  $(21.73)$ &   \\
0.97-1.00 &        602 &        253 &        150 &       1005 \\
\       &   $(6.71)$ &  $(18.42)$ &  $(32.07)$ &   \\
1.00-1.03 &        744 &        455 &        237 &       1436 \\
\       &   $(16.69)$ &  $(29.77)$ &  $(41.62)$ &  \\
1.03-1.06 &        800 &        505 &        180 &       1485 \\
\       &   $(30.28)$ &   $(41.4)$ &  $(53.41)$ &  \\
$\geq 1.06$ &       1862 &        827 &        125 &       2814 \\
\       &   $(62.18)$ &  $(64.09)$ &  $(66.16)$ &  \\
\midrule
Subtotal &   4256 &       2271 &        793 &       7320 \\
\bottomrule
\end{tabular}}
\label{tab: outofsampledata}
}
\caption{Sample properties of the in-sample (sample A) and the out-of-sample (sample B) option data, reported by dividing the data into three groups based on maturity (trading days) and six groups based on moneyness (P/K). The table shows the number of options and the average price (in parentheses).} 
\label{TAB:Sample_A&B}   
}
\end{table}

One approach is to estimate the structural parameters and spot volatility on each trading day separately \citep*[see, e.g.,][]{Bakshi97} and another to use more than one day of option prices to estimate the structural parameters \citep*[see, e.g.,][and references therein]{Christoffersen04}. There are numerous reasons for preferring a multi-day sample. First and most importantly, it is essential that we use information on time-variation in the price of the underlying asset and spot volatility; this can be done only with multi-day data. Second, the first approach (single-day samples) would yield different estimates of the structural parameters for each day while, on the other hand, structural parameters are assumed constant. 
Third, \citet{Cont04} argue that  ``given that the number of calibration constraints (option prices) is finite (and not very large), there may be many L\'evy triplets which reproduce call prices with equal precision.'' Overcoming, or at least minimizing, the instability problem was enough incentive for us to increase the number of option price observations using multi-date data. Fourth, assessing long-term performance with single-day samples is problematic.

Overall, it is better to use more than one day of option prices to estimate the structural parameters. In the present literature, such multi-day estimation have been applied with GARCH type models \citep*[see, e.g.,][]{Heston00,Christoffersen04}. Then a volatility updating rule can be used to link volatility on different dates for given structural parameters \citep*[see][Online Appendix]{Christoffersen04}. However, this is not possible with an option pricing model, in which the stock return process and the instantaneous volatility process are driven by separate random terms. Consequently, we need a suitable proxy for market volatility, which is intrinsically unobservable. 
The recent empirical literature has constructed volatility proxies from volatility indices, like VXO and VIX. On September 22, 2003, the CBOE reformulated its implied volatility index to use the model-free implied volatility approach on S\&P 500 and created a historical record for the changed  S\&P 500 VIX dating back to 1990.\footnote{See CBOE Documentation 2003.} This reformulated VIX is used as a spot volatility proxy in, e.g., \citep{Jones03,Bakshi06,Ait-Sahalia07,Christoffersen11}. The Black-Scholes implied volatility could be adjusted for the effect of mean revision in volatility, but as \citet{Ait-Sahalia07} show, with VIX, this adjustment has only a marginal effect on the results, perhaps because the current reformulated VIX is model-free. Therefore, as in \citep{Bakshi06} and also partly in \citep{Ait-Sahalia07}, we use the reformulated model-free VIX data directly as a proxy for instantaneous volatility. To be precise, we use the lagged values such that  yesterday's closing value of VIX serves as a volatility proxy for today's model-based option prices. Notice that when pricing options, volatility is modeled under the risk-neutral measure, and VIX measures volatility just under the risk-neutral measure.

The loss function we use is the square root dollar mean-squared error,
\[
\$\text{RMSE} = \sqrt{\frac{1}{n} \sum_{i}^n \left(\hat{c}(t_i, P_{t_i}, T_i, K_i) - c(t_i, P_{t_i}, x_{t_i}, T_i, K_i, r^{\text{T-bill}}_i, \bar{\alpha}; \theta) \right)^2},
\]
which is minimized with respect to the structural parameters, $\theta$. Here $n$ denotes the number of contracts and $r^{\text{T-bill}}_i$ the observed T-Bill rate. Moreover, $\hat{c}(\cdot)$ is the price of the $i$-th option, $c(\cdot;  \theta)$ the corresponding model price, {  and} $x_{t_i} \equiv \text{VIX}_{t_{i}-1}$ denotes the volatility proxy used at time $t_i$. 

The loss function was minimized using the Nelder-Mead simplex algorithm, a derivative-free methods for unconstrained multivariable function minimization, as implemented in the \textsc{Matlab} \texttt{fminsearch} code \citet{OptimizationToolbox10}. The same optimization algorithm has been used with Monte Carlo simulations at least in \citep{Barone08}. The options are priced by simulating 20,000 paths with antithetic variates. To speed up computation, distributed computing has been used such that workers (cores) calculate option prices for different days independently; i.e., worker one calculates the options for day one, worker two for day two, and so on. This {  requires 150 workers and achieves a 150-fold speedup}.

\subsection{Results}

Table \ref{TAB:results} shows  the evaluation results for three specifications: (i) full model; (ii) constant $\gamma=0$; (iii) constant $\gamma=0$ with {  doubled} interest rates and {  halved} dividend growth rate. Before presenting and interpreting the results, we emphasize that our estimates concern the volatility process (\ref{EQ:dx}), not directly with respect to the squared volatility process (\ref{EQ:dh}), though they can be converted by applying the relations $\kappa = 2\beta$, $\sigma_h = 2 \sigma_x$, and $\theta = \sigma_x^2/(2 \beta)$. Moreover, the volatility risk premium can be presented with respect to squared volatility process as $\lambda_h = \tilde{\kappa} - \kappa$, whereas with respect to volatility process we express it as $\lambda_x = \tilde{\beta} - \beta$. Clearly, our result can be converted with $\lambda_h = 2 \lambda_x$. Therefore, when the estimated values of $\beta$, $\sigma_x$, and $\lambda_x$ are compared with the estimates of $\kappa$, $\sigma_h$, and $\lambda_h$ in the early literature, our estimates should be multiplied by two.

\begin{table}[h]
\linespread{1} 
\begin{center}
\footnotesize{
\begin{tabular}{lccc}

           & \multicolumn{ 1}{c}{(i) $\gamma$ free} & \multicolumn{ 1}{c}{(ii) $\gamma = 0$} & (iii) $\gamma = 0$ with \\

           & \multicolumn{ 1}{c}{} & \multicolumn{ 1}{c}{} & $r_i = 2 r^{\text{T-bill}}_i$, $\alpha = \bar{\alpha} / 2$ \\
\hline 
\\

$\tilde{\beta}$ &     1.2476 &    32.8570 &     1.3282 \\

           &   (0.0420) &   (2.5897) &   (0.0474) \\

$\sigma_x$ &     0.2713 &     0.6659 &     0.2666 \\

           &   (0.0032) &   (0.0271) &   (0.0035) \\

$\rho_{dx}$ &    -0.6410 &     0.4241 &    -0.8002 \\

           &   (0.0069) &   (0.0180) &   (0.0033) \\

$\lambda_x$ &    -0.3376 &            &            \\

           &   (0.0128) &            &            \\

  $\gamma$ &     1.7929 & 					 & 					 \\

    {\bf } &   (0.0099) &            &            \\
\hline
   \$RMSEs &            &            &            \\

  Sample A &     0.8111 &     1.1327 &     0.9114 \\

  Sample B &     0.9355 &     1.6429 &     0.9859 \\

\end{tabular}  
}
\caption{We estimate our model directly by fitting the observed option prices using a nonlinear least-squares {  code} to minimize \$RMSE. Only options in Sample A (June 3 - August 4, 1995), consisting of 11,276 contracts, are used in the estimation. Standard errors are reported below each parameter estimate in parentheses. At the bottom, the table reports \$RMSE for samples A and B at the parameter optima. Results are reported for three specifications. Specification (i) represents a general model in which $\gamma$ is a free parameter. Specification (ii) comes with $\gamma=0$ (no risk-return trade-off) as also does specification (iii), but the observed average dividend growth rate, $\bar{\alpha}$, and T-bill rates, $r^{\text{T-bill}}$, have been adjusted. In particular, we use {  doubled} T-bill rates and  {  halved} dividend growth rate in (iii).}
\label{TAB:results} 
\end{center}
\end{table}
For specification (i), we find that the estimates of $\tilde{\beta}$ and $\sigma_x$ are reasonable and {  fairly} consistent with the early empirical results of the Heston (1993) model. Moreover, the correlation between return volatility and dividend growth rate is negative, indicating the existence of a leverage effect. All the parameters are statistically significant. More importantly, the estimated $\gamma$ is about 1.8 and statistically significant. This estimate implies that the expected returns are related to squared return volatility and that, consequently, the price-dividend ratio is sensitive to return volatility. Our estimate is thus consistent with the ICAPM and the theory of volatility feedback. Recently, after a 20-year debate, studies have, after developing estimation approaches, identified positive and significant relations between the expected rate of return and volatility \citep*[see, e.g.,][]{Ghysels05,Bali06,Bali08,Bekaert09,Bali10} and the references therein), and our results provide further evidence of this significant and positive relation using an entirely novel approach. Moreover, we find that the estimate of $\lambda_x$ is negative (-0.338), and that it has been estimated accurately as well. In addition, this estimate is consistent with {  published} theory and findings \citep*[see][]{Pan02,Bakshi03,Carr09,Bollerslev11a}. Based on the difference between option-implied and realized volatilities, \citet{Bollerslev11a} construct a volatility risk premium index and show also a link between volatility risk premium and the price of diffusion return risk (the risk-return trade-off coefficient). Following \citet{Heston93}, \citet{Bakshi03}, and \citet{Bollerslev11a}, we can, in line with our assumptions, write that
 \[
 \lambda(x_t) = \gamma \frac{\ud }{\ud  \tau}\mathrm{Cov}_t(x_t, R_t)|_{\tau = t},
 \]
 which yields with $\lambda(x) = \lambda_x(x) x$ that
\[\begin{split}
   \lambda_x(x_t) &= \gamma \sigma_x \frac{\ud }{\ud \tau} \mathrm{Corr}_t(x_t, R_t)|_{\tau = t}\\
   &= \gamma \sigma_x \rho_{rx}(x_t),
\end{split}\]
where $\rho_{rx}(x)$ is determined according to Eq.\ (\ref{EQ:rho_rx}). Hence a positive price of diffusion return risk and a negative correlation between the returns and return volatility together imply a volatility risk premium. Because the standard Heston model assumes that the correlation between the return and return volatility is constant, then also the volatility risk premium is constant in time. Under our transversality settings, the volatility risk premium becomes dependent on return volatility and is hence time-varying and stochastic, yet very stable with respect to $x$. With our estimates, $\rho_{rx}$ evolves around $-0.7879$ and using this value together with $\sigma_x = 0.27$ and $\gamma = 1.79$, we obtain an implied volatility risk premium of about $-0.38$. Because this is close to our unrestricted estimate of $\lambda_x$, $-0.338$, the above theoretical relation between the price of diffusion return risk and the volatility risk premium seems to hold in the light of our data. Compared to the other studies, the in-sample and out-of-sample RMSEs are reasonable low \citep*[see, e.g.,][]{Christoffersen04,Christoffersen08b}.

Specification (ii) assumes that $\gamma=0$, excluding volatility feedback and implying that the price-dividend ratio, and then also the dividend yield, do not depend on return volatility. As explained earlier, $1/f = r - \alpha$ is constant and not dependent on $\beta$; hence the volatility risk premium cannot be estimated. Both the in-sample (sample A) and out-of-the-sample (sample B) RMSEs are substantially higher than in specification (i). We immediately notice that the estimate of $\beta$ is not reasonable, for it implies that $\kappa$ is greater than 60, which is absolutely unrealistic. Moreover, the correlation between return volatility and dividend growth is substantially positive and contradicts the leverage effect. On the other hand, we observe that the assumption of $\gamma = 0$ is not consistent with the observed interest rates and the dividend growth rate. In particular, the average dividend growth rate, $\bar{\alpha} \approx 0.0613$, exceeds the T-bill rates that range from 0.054 to 0.0596 and imply that the dividend yield, $1/f = r - \alpha$, takes negative values and does not satisfy the assumption of positive dividend yields. To put it simply, in 1995, the dividend growth rate was greater than the risk-free interest rates, and this phenomenon cannot be captured with $\gamma=0$. Therefore, we constructed a third specification, by which we ensured that the dividend yield is always a positive constant. 

In specification (iii), we use $r_i = 2 r^{\text{T-bill}}_i$ with $\alpha = \bar{\alpha} / 2$, where $r^{\text{T-bill}}_i$ is the T-bill rate at date $i$. That is, to have strictly positive dividend yields with $\gamma=0$, we double the observed T-bill rates and halve the observed average dividend growth rate. As can be seen from Table \ref{TAB:results}, this adjustment makes parameter estimates with $\gamma=0$ more realistic. Furthermore, both the in-sample and out-of-the-sample RMSEs decrease yet remain greater than in specification (i). We emphasize here that in any interpretation of the estimates of specification (iii), it should be kept in the mind that these adjusted interest rate and dividend growth rate values do not represent the true economic situation in 1995. Rather, this specification shows that the poor performance and unrealistic estimates of specification (ii) can be partly explained by the fact that the assumption of $\gamma=0$ contradicts the empirical observations of dividend growth and risk-free interest rates.

\section{Conclusion}

This paper shows that under the volatility feedback effect the stock price process takes a form in which the dividend yield and the correlation between returns and return volatility become time-varying and endogenously determined by total return volatility, implying that the stock price and dividend stream do not always move in the same direction and that the ratio of total return volatility to dividend growth volatility can be relatively high. The main implication is that in contrast to the current wisdom, there is a {  mechanism by} which the market price of diffusion return risk (or equity risk-premium) affects option prices, i.e. the market price of return risk is needed as an input to price options under our framework. Moreover, we demonstrate how the price of a call option can be decreasing in squared total return volatility. As a part of the calibration progress, we present an approach to identifying the risk-return relation using forward-looking option data. We also show how a positive risk-return relation agrees with empirical observations of dividend growth and risk-free interest rates. Overall, we argue that the prevailing practice of ignoring the time-varying price-dividend ratio in option pricing oversimplifies the stock market dynamics and {  disagrees} with data and theory.

This framework has potential for further research. In particular, one could examine the hedging of volatility risk under volatility feedback, a particularly interesting topic, because, as we show, depending on the moneyness, option prices can react positively or negatively to an increase in squared return volatility. Furthermore, it would be worth considering, within our framework, a stochastic investment opportunity with an intertemporal hedging component to hedge against changes in the forecasts of future market volatilities. Such an experiment is motivated by recent empirical evidence of a significantly negative relation between expected return and volatility risk \citep*[see][]{Bali10}. In this paper, we aimed to illustrate how to obtain option-implied forward-looking values for the price of diffusion return risk and volatility risk-premium without the need of time-series stock data, but in future research, it would be interesting to investigate the volatility feedback using data on options together with stock price and dividend series. On the other hand, direct estimation based on time-series data would be challenging at best. Moreover, now that we have focused {  on} stock price and dividend dynamics, option prices, and volatility feedback with a diffusion-based volatility model, our extension could be addressed also with more elaborate volatility models with jumps or non-affine stochastic volatility models. These general suggestions for future research may, however, require increasing the number of model parameters and computational complexity, which would again challenge especially empirical analysis. Therefore, we expect work to continue also on complementary computational solutions to ease our progress.

\section*{Technical Appendix: Numerical solution of boundary value problem}

The ordinary differential equation boundary value problem (BVP) in Proposition~\ref{PROP:DE} can be solved numerically using general-purpose numerical solvers. We used the \texttt{bvp4c} solver 
in the  \textsc{Matlab} interactive scientific computing software system. The \texttt{bvp4c} solver uses a fourth-order collocation method with automatic mesh refinement.
The solver's algorithms are described in \citet{bvp4c}; here we provide details of the formulation of the problem for numerical solution.

For numerical solution the BVP solution domain $x\in[0,\infty)$ is approximated by a finite interval $[0,b]$. We used $b=5$; this is considered adequate because repeating the computations with larger $b$ values gave the same solution to within 4  decimals. Empirically, return volatility can hardly be greater than 100\%; therefore $b=5$ is also empirically reasonable.

For large $x$, the dominant term on the left hand side of the differential equation~(\ref{EQ:DE}) is $-\gamma x^2 f(x)$. Balancing this term with the right hand side leads to the asymptotic estimate $ f(x)= \frac{1}{\gamma x^2}$. On the basis of this asymptotic analysis, we used the boundary condition $ f(b)= \frac{1}{\gamma b^2}$ in the numerical solution.
 The  symmetry condition $f_x(0)=0$ was used as the second boundary condition. 
 
The \texttt{bvp4c} solver supports the use of parametric continuation whereby the problem is successively solved with gradually changing parameter values. We exploit this feature by first solving the problem with $\rho_{dx}=0$, then successively solving the problem with $\rho_{dx}$ values gradually increasing to the desired level. Because the BVP with $\rho_{dx}=0$ is linear,
the numerical solution of the first problem can be obtained directly without any Newton iterations. As the $\rho_{dx}$ values are increased, the solution and mesh from the previous solution are used as initial estimates for the Newton process, which then typically converges in a few iterations.

\bibliographystyle{model1-num-names}

\bibliography{references}

\newpage







\end{document}